\documentclass[10pt]{article}
\usepackage[a4paper,left=3cm,right=3cm,top=3cm,bottom=3cm]{geometry}
\usepackage{setspace}
\usepackage[utf8]{inputenc}

\usepackage{bm}

\usepackage[style=chem-acs, backend=biber, doi=false,articletitle=true]{biblatex} 
\usepackage{todonotes}

\usepackage{csquotes}

\usepackage{hyperref}

\usepackage{graphicx}
\usepackage{float}
\usepackage{booktabs}
\usepackage{multirow}
\usepackage{subfigure}
\usepackage[labelfont=bf, font={small}]{caption}
\usepackage{chngcntr}
\usepackage{diagbox}
\usepackage{slashbox}
\usepackage{cellspace}
\usepackage{chemfig}
\makeatletter
\def\Hv@scale{.95}
\makeatother
\DeclareMathAlphabet{\foo}{OT1}{phv}{m}{n}

\usepackage{amsbsy,amsmath,amssymb}

\usepackage{chemformula}

\def\d{{\mathrm{d}}}

\def\fde{FDE}%{\texttt{FDE}}}
\def\pe{PE}%{\texttt{PE}}}
\def\nopol{{\texttt{NOPOL}}}
\def\gspol{{\texttt{GSPOL}}}
\def\dpol{{\texttt{DPOL}}}
\def\eef{{\texttt{DPOL+EEF}}}
\def\pna{{\textit{p}NA}}
\def\ie{{\textit{i.e.}}}
\def\via{{\textit{via}}}
\def\dz{{aug-cc-pVDZ}}%double-$\zeta$}}
\def\tz{{aug-cc-pVTZ}}%triple-$\zeta$}}
\def\s{{$\mathcal{S}$}}
\def\f{{$\mathcal{F}$}}
\def\fshift{{$\mathcal{F}$-shift}}
\def\fshifts{{$\mathcal{F}$-shifts}}

\usepackage{pifont}
\newcommand{\cmark}{\ding{51}}%
\newcommand{\xmark}{\ding{55}}%
\usepackage[floats]{preview}
\PreviewEnvironment{tikzpicture}
\usepackage{pgfplots}
\usepackage{tikz}
\usetikzlibrary{shapes.geometric, arrows,fit, shadows}
\pgfplotsset{compat=1.16} 
\tikzstyle{every picture}+=[font=\sffamily]
\pgfdeclarelayer{background}
\pgfdeclarelayer{foreground}
\pgfsetlayers{background,main,foreground}
\definecolor{white}{RGB}{255,255,255}% white
\definecolor{luhblue}{RGB}{0,80,155}% university dark blue
\definecolor{luhblue40}{RGB}{153,185,216}% university dark blue 40% transparency
\definecolor{luhblue20}{RGB}{204,220,235}% university dark blue 20% transparency
\definecolor{luhblack}{RGB}{0,0,0}% 
\definecolor{luhblack40}{RGB}{153,153,153}% 
\definecolor{luhblack20}{RGB}{204,204,204}% 
\definecolor{luhgreen}{RGB}{200,211,23}% university

\begin{document}

\title{Theoretical and numerical comparison of quantum- and classical embedding models  for optical spectra}
\author{Marina Jansen, Peter Reinholdt, Erik D.\ Hedeg{\aa}rd, and Carolin K\"onig}
\maketitle
%%%%%%%%%%%%%%%%%%%%%%%%%%%%%%%%%%%%%%%%%%%%%%%%
\section*{Abstract \label{sec:abstract}}
%%%%%%%%%%%%%%%%%%%%%%%%%%%%%%%%%%%%%%%%%%%%%%%%
Quantum-mechanical (QM) and classical embedding models approximate a supermolecular quantum-chemical calculation. 
This is particularly useful when the supermolecular calculation has a size that is out of reach for present QM models. 
Although QM and classical embedding methods share the same goal, they approach this goal from different starting points. 
In this study, we compare the polarizable embedding (PE) and frozen-density embedding (FDE) models. 
The former is a classical embedding model, whereas the latter is a density-based QM embedding model. 

Our comparison focuses on solvent effects on optical spectra of solutes. 
This is a typical scenario where super-system calculations including the solvent environment become prohibitively large.    
We formulate a common theoretical framework for PE and FDE models and systematically investigate how PE and FDE approximate solvent effects. 
Generally, differences are found to be small, except in cases where electron spill-out becomes problematic in the classical frameworks. 
In these cases, however, atomic pseudopotentials can reduce the electron-spill-out issue. 

%%%%%%%%%%%%%%%%%%%%%%%%%%%%%%%%%%%%%%%%%%%%%%%%
\section{Introduction \label{sec:intro}}
%%%%%%%%%%%%%%%%%%%%%%%%%%%%%%%%%%%%%%%%%%%%%%%%
Quantum-mechanical (QM) methods are indispensable for the calculation of optical spectra, but their use often becomes computationally too demanding for large systems. 
Embedding schemes have been introduced to circumvent the full, super-system QM calculation by including large environments  through an effective \textit{embedding} operator. \\
The definition of an embedding model requires that the system is split into an active system and the remaining part ("the environment").\cite{Warshel1976, Singh1986, Field1990, Senn2009}
Embedding approaches can be divided into two main classes:
(i) QM-classical embedding approaches describe the active system by a QM method, whereas all interactions between the active system and environment (as well as the environment itself) 
  are treated by a classical description. (ii) QM-QM embedding  describes both active system and environment with QM methods 
 (either on the same or different footings). In this case, the interaction between active system and the environment also contains QM contributions.
\\
For optical properties, the electrostatic interaction between  the active system and the environment is often 
 the dominating embedding contribution. 
In traditional QM-classical approaches, this contribution is modeled through (atomic) point charges in the environment.\cite{Senn2009}
The point-charge model is, however, insufficient in many cases.\cite{Warshel1976, Thompson1996, Curutchet2009, Defusco2011, Daday2013, Boulanger2018} 
Therefore, a large number of more advanced embedding schemes have been developed over the years.\cite{Thompson1995, 
Gao1997, Lin2007, Defusco2011,List2016a,Bondanza2020,Loco2021,Lipparini2021,bookchapter, Jensen2003a,Jensen2003b,Olsen2010,Thompson1996,Gordon2007,Gordon2013,Loco2016} 
\\
In this work, we employ an advanced QM-classical embedding model, namely the polarizable embedding (PE) model\cite{Olsen2010}. 
In this model, point charges in the environment are replaced by a multipole expansion. 
Additionally, PE incorporates the environment polarization through anisotropic electronic dipole--dipole polarizabilities. 
The parameters for the environment (\ie{}~multipoles and polarizabilities) are obtained from QM calculations on isolated fragments. If the environment is a solvent, these fragments are most naturally defined as solvent molecules.
\\
In the class of QM-QM embedding methods, the total system is expressed by means of 
 fragments or \textit{subsystems} \cite{Yang1991, Senatore1986a, Johnson1987, Cortona1991a}: 
Within density functional theory (DFT) this is known as subsystem DFT.\cite{Jacob2014}
All subsystems are described by their electron densities, which are obtained by quantum-chemical calculations. 
The interaction of the environment subsystems with the active subsystem is then recovered through an embedding potential, which contains quantum-mechanical contributions. 
This potential is dependent on all other subsystem's electron densities.\cite{Wesolowski1993, Jacob2014}
In practice, the environmental electron densities are commonly kept frozen, so that only the active subsystem's electron density is polarized. 
In this frozen-density embedding (FDE) approach\cite{Wesolowski1993}, the effect of the environment's polarizability can be incorporated by 
 self-consistently cycling through all subsystems in a  so-called freeze-and-thaw scheme.\cite{Senatore1986a, Cortona1991a, Wesolowski1993, Laricchia2011, Jacob2014}
\\
Both embedding classes share, hence, a common goal and have been employed to show that po\-la\-ri\-za\-tion effects originating from the environment can play a significant role in the accurate calculation of local optical properties.\cite{Filippi2012, Daday2013, Daday2015, Schwabe2015, Olsen2011b}  Yet, direct numerical comparisons have been rare due to their different formulations and implementations.\cite{Jacob2006, Gomes2012} %Mention examples of studies and review.
We recently developed a common theoretical framework\cite{bookchapter} encompassing both fragmentation-based QM-QM and QM-classical embedding methods with a special focus on FDE and PE. 
This framework was employed to dissect  how the two classes of embedding models describe the interactions between the active system and the environment. We here continue this comparison by quantifying how the theoretical differences manifest numerically for optical properties of two solvated systems, employing a supermolecular  calculation as a reference.

\begin{figure}
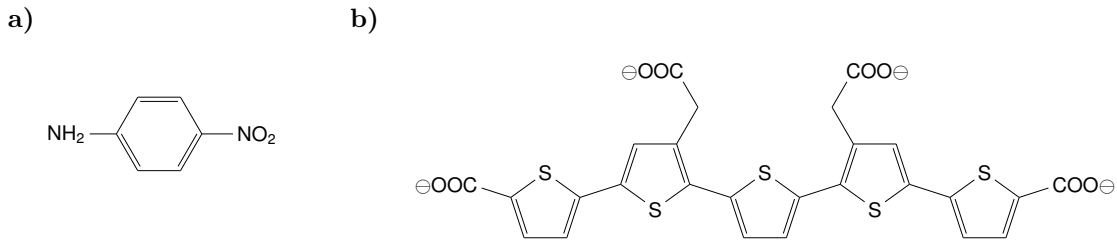
%[H]
    \centering
    \begin{minipage}[t]{0.25\textwidth}
    \textbf{a)}\vfill
    \vspace{8mm}
            \scalebox{.8}{\hspace{15pt}
\chemfig[atom sep=2em]{[:90]*6(=(-NO_2)-=-(-NH_2)=-)}
}
\vfill
    \end{minipage}\hfill
   \begin{minipage}[t]{0.7\textwidth} 
    \textbf{b)}\vfill
    \begin{center}
    \scalebox{.8}{\hspace{15pt}
\chemfig[atom sep=2em]{\ominus OOC-[:-18]*5(=-=(-(*5(-S-(-(*5(=-=(-(*5(-S-(-(*5(=-=(-COO\ominus)-S-)))=-(-[:120]-[:60]COO\ominus)=)))-S-)))=(-[:60]-[:120]\ominus OOC)-=)))-S-)}
}
    \end{center}
    \end{minipage}%
%    \vspace*{5mm}
    \caption{Dyes considered in this work: (a) \textit{para}-nitroaniline (b) pentameric formyl thiophene acetic acid 
    \label{fig:molecules}}
\end{figure}

Our target systems (fig.~\ref{fig:molecules}) are two fluorescent dyes whose excited-state properties are known to be sensitive to solvent effects: 
The first target system is \textit{para}-nitroaniline (\pna{}), which has been studied with several different embedding schemes \cite{Fujisawa2006, Kosenkov2011, Slipchenko2010, Sneskov2011, Defusco2011, Daday2013}.
Yet, the performance of the QM-QM and QM-classical embedding schemes have never been  compared in a combined  study. The second test case is pentameric formyl thiophene acetic acid (pFTAA), a luminescent biomarker developed 
 for fluorescence imaging for amyloid proteins.\cite{Klingstedt2015}
The mechanism occurring with the chromophore embedded in the protein site is not fully understood yet, but it is known that  its properties strongly depend on the solvent--solute interactions 
 and the conformation of the molecule.\cite{Sjoqvist2011, Sjoqvist2014, Gustafsson2020, Gustafsson2021}
Notably, pFTAA is an anionic system and therefore poses somewhat different challenges in the description of the solute--solvent interaction than \pna{}.
\\

This paper is organized as follows.
We first briefly introduce the PE and FDE scheme in the common theoretical framework as derived in previous work\cite{bookchapter}  (sec.~\ref{sec:theory}).
In particular, we point out similarities and differences between these schemes.
We describe the computational setup, that was developed to enable us to compare the embedding methods on equal footing  (sec.~\ref{sec:implementation}). 
In section~\ref{sec:comp_details}, the computational details are given.
Subsequently, the results are presented and discussed (sec.~\ref{sec:results}). 
Finally, in the last section (sec.~\ref{sec:conclusion}), we conclude and summarize the study.
\\

%%%%%%%%%%%%%%%%%%%%%%%%%%%%%%%%%%%%%%%%%%%%%%%%
\section{Theoretical Background\label{sec:theory}}
%%%%%%%%%%%%%%%%%%%%%%%%%%%%%%%%%%%%%%%%%%%%%%%%
In this section, we give a brief overview of the density-based QM-QM embedding and classical PE methods. For a more detailed derivation, reviews, and further extensions of the presented models we refer to Refs.~\cite{bookchapter, Bondanza2020, Loco2021, Lipparini2021, Wesolowski2015, Jacob2014, Krishtal2015, Sun2016, Gomes2012}. Here, we use the common framework developed in Ref.~\cite{bookchapter}.

Embedding schemes commonly focus on a selected, \textit{active} subsystem A. The remaining parts are labelled \textit{the environment} (env). 
The total energy can be written as the sum of the energy functionals of the respective electron densities $\rho_{\mathrm{A}}$ and $\rho_{\mathrm{env}}$,
\begin{align} 
	E_\mathrm{tot}[\rho_\mathrm{tot}] = E_\mathrm{A}[\rho_{\mathrm{A}}] + E_\mathrm{env}[\rho_{\mathrm{env}}] + E_\text{int}[\rho_{\mathrm{A}},\rho_{\mathrm{env}}],
	\label{eq:int_total_general}
\end{align} 
where  $E_{\mathrm{tot}}[\rho_\mathrm{tot}]$ and $E_{\mathrm{A}}[\rho_\mathrm{A}]$ denote 
 the total energy functional and the energy functional of the active subsystem A, respectively. The
 $E_\text{int}[\rho_{\mathrm{A}},\rho_{\mathrm{env}}]$ term  defines the interaction energy, \ie{} 
$E_\mathrm{int}[\rho_{\mathrm{A}},\rho_{\mathrm{env}}] = E_\mathrm{tot}[\rho_\text{tot}] - E_\mathrm{A}[\rho_{\mathrm{A}}] - E_\mathrm{env}[\rho_{\mathrm{env}}]$, 
 where we assume $\rho_\text{tot}=\rho_{\mathrm{A}}+ \rho_{\mathrm{env}}$. 
The environment density can analogously be approximated as a sum of all environmental (subsystem) densities, $\rho_{X}$,  by $\rho_{\mathrm{env}}= \sum_{X\neq \mathrm{A}}\rho_{X}$. 
We further decompose the interaction energy into Coulomb (classical), $E_\text{int}^\text{C}[\rho_{\mathrm{A}},\rho_{\mathrm{env}}]$, and QM contributions, $E_\text{int}^\text{QM}[\rho_{\mathrm{A}},\rho_{\mathrm{env}}]$,
 \begin{align}
 	E_\text{int}[\rho_{\mathrm{A}},\rho_{\mathrm{env}}] = 
 	E_\text{int}^\text{C}[\rho_{\mathrm{A}},\rho_{\mathrm{env}}]
 + 	E_\text{int}^\text{QM}[\rho_{\mathrm{A}},\rho_{\mathrm{env}}], 
 	\label{eq:EintQMandenv}
 \end{align} 
where both contributions can be obtained analogously to eq.~\eqref{eq:int_total_general}. 

The Coulomb part is included in both density-based and classical embedding schemes. 
In the density-based QM-QM embedding schemes this part is expressed as 
 \begin{align}
 	E_{\mathrm{int}}^{\mathrm{C}}[\rho_{\mathrm{A}},\rho_{\mathrm{env}}]  =&  
 	\sum_{X\neq \mathrm{A}} \int \int \frac{\rho_{\mathrm{A}} (\mathbf{r}_\text{a}) \rho_{X}(\mathbf{r}_x) }{|\mathbf{r}_\text{a} - \mathbf{r}_x |} \d \mathbf{r}_\text{a}\d \mathbf{r}_x 
 	- \sum_{X\neq \mathrm{A}} \sum_{I \in \mathrm{A}} \int \frac{Z_I \rho_{X}(\mathbf{r}_x)}{|\mathbf{R}_I - \mathbf{r}_x|} \d \mathbf{r}_x \notag \\
 	& - \sum_{X\neq \mathrm{A}}  \sum_{J \in X} \int \frac{Z_J \rho_{\mathrm{A}}(\mathbf{r}_\text{a})}{| \mathbf{R}_J - \mathbf{r}_\text{a}|} \d \mathbf{r}_\text{a} 
 	 +  \sum_{X\neq \mathrm{A}}  \sum_{I \in \mathrm{A}} \sum_{J \in X} \frac{Z_I Z_J}{|\mathbf{R}_I - \mathbf{R}_J|}, 
 	\label{eq:e_coul}
\end{align}
where $\mathbf{R}_{I/J}$ denote nuclear coordinates, $ \mathbf{r}_{a/x}$ denote electronic coordinates, and  $Z_{I/J}$ are nuclear charges.  
The interaction of the environment with the active subsystem is included \textit{via} Coulomb [and possibly quantum-mechanical (QM)] contributions through an effective embedding operator.
The effective Hamiltonian for subsystem A can be expressed as
\begin{align}
\hat{H}_\mathrm{A}^{\text{eff}}  
= \hat{H}_{\mathrm{A}} + \hat{v}_{\text{A}}^\mathrm{emb}  
= \hat{H}_{\mathrm{A}} + \int\hat{\rho}_\mathrm{A}(\mathbf{r}_\mathrm{a}) {v}_{\text{A}}^\mathrm{emb}(\mathbf{r}_\mathrm{a}) \mathrm{d}\mathbf{r}_{\mathrm{a}},
\label{eq:heff}
\end{align}
where the density operator $\hat{\rho}_{\mathrm{A}} (\mathbf{r}_\text{a}) = \sum_{i \in \mathrm{A}}\delta (\mathbf{r}_i - \mathbf{r}_\text{a} )$
defines the connection between the Hamiltonian with electron-based coordinates and density-based expressions with real-space coordinates.
We define the embedding operator via a real-space potential,
\begin{align}
v_{\mathrm{A}}^\text{emb} (\mathbf{r}_\mathrm{a}) 
= &
\frac{\delta}{\delta\rho_\mathrm{A}} 
(
E_\mathrm{tot}[\rho_\mathrm{tot}] 
- E_\mathrm{A}[\rho_\mathrm{A}]
) \notag \\
%= &
%\frac{\delta}{\delta\rho_\mathrm{A}} 
%(
%E_\mathrm{env}[\rho_\mathrm{env}]
%+ E_\mathrm{int}[\rho_\mathrm{A}, \rho_\mathrm{env}]
%) \\
= &
\frac{\delta  E_\mathrm{tot}^\mathrm{C}[\rho_\mathrm{tot}]  }{\delta \rho_{\mathrm{A}}} 
- \frac{\delta  E^\mathrm{C}_\mathrm{A}[\rho_\mathrm{A}]  }{\delta \rho_{\mathrm{A}}} 
+ \frac{\delta  E_\mathrm{tot}^\mathrm{QM}[\rho_\mathrm{tot}]  }{\delta \rho_{\mathrm{A}}} 
- \frac{\delta  E^\mathrm{QM}_\mathrm{A}[\rho_\mathrm{A}]  }{\delta \rho_{\mathrm{A}}}. 
\label{eq:vemb}
\end{align}  
In FDE, the environmental density $\rho_\text{env}$ is kept frozen so that the total embedding potential resulting from eq.~\eqref{eq:vemb} becomes
\begin{align}
v_\mathrm{A}^\mathrm{FDE} (\mathbf{r}_a) 
%= &
%\left.  \frac{\delta  E_\mathrm{int}^\mathrm{C}[\rho_{\mathrm{A}},\rho_{\mathrm{env}}]}{\delta \rho_{\mathrm{A}}} \right|_{\rho_{\mathrm{env}}=\text{const.}}
%    + \left. \frac{\delta  E^{\mathrm{QM}}_\mathrm{int}[\rho_{\mathrm{A}},\rho_{\mathrm{env}}]}{\delta \rho_{\mathrm{A}}}\right|_{\rho_{\mathrm{env}}=\text{const.}} 
%\\
= &
v_{\mathrm{A}}^\mathrm{C (FDE)} [\rho_{\mathrm{env}}] (\mathbf{r}_{\mathrm{a}}) 
+  
v_\text{A}^\text{nadd,kin}[\rho_{\mathrm{A}},\rho_{\mathrm{env}}] (\mathbf{r}_{\mathrm{a}})
+  
v_\text{A}^\text{nadd,xc}[\rho_{\mathrm{A}},\rho_{\mathrm{env}}] (\mathbf{r}_{\mathrm{a}}), 
\label{eq:FDEpot}
\end{align}
with the Coulomb potential only depending on $\rho_{A}$ and the (frozen) densities of the environment %\erik{what parts are important to connect this equation with the one below?}
\begin{align}
	v_{\mathrm{A}}^\mathrm{C (FDE)} [\rho_{\mathrm{env}}] (\mathbf{r}_{\mathrm{a}})
& = 
\frac{\delta  E_\mathrm{int}^\mathrm{C}[\rho_\mathrm{A}, \rho_\mathrm{env}]  }{\delta \rho_{\mathrm{A}}}  \notag \\
& = 
	- 
	\sum_{X\neq \mathrm{A}}\sum_{J \in X} 
	\frac{Z_{J}}{|\mathbf{r}_{\mathrm{a}} - \mathbf{R}_{J} | }
	+
	\sum_{X\neq \mathrm{A}}\int \frac{\rho_X (\mathbf{r}_{x})}{|\mathbf{r}_{\mathrm{a}} - \mathbf{r}_{x} | }
	\d \mathbf{r}_{x},
\label{eq:emb_rho}
\end{align}
where $ E_\mathrm{int}^\mathrm{C}[\rho_\mathrm{A}, \rho_\mathrm{env}]$ is defined in eq.~\eqref{eq:e_coul}.
The QM contributions from eq.~\eqref{eq:vemb} are comprised of  kinetic and an exchange--correlation (xc) parts, represented by $v_\text{A}^\text{nadd,kin}$ and
  $v_\text{A}^\text{nadd,xc}$ in eq.~\eqref{eq:FDEpot}.   
In practical calculations, these contributions are often approximated by orbital-free DFT methodologies\cite{Wesolowski1993, Jacob2014, Krishtal2015}, though for $v_\text{A}^\text{nadd,kin}$ also orbital-dependent projection schemes have been reported.\cite{Goodpaster2010,Manby2012,Khait2012,Chulhai2015,Hegely2016,Chulhai2017,Culpitt2017,Ding2017,Lee2019,Graham2020,Graham2022,Bensberg2019,Bensberg2019a}
\\
The application of a fixed $\rho_{\mathrm{env}}$  
 in FDE leads to several possible choices of frozen densities. The crudest approximation is to use the density from the isolated fragments by a superscript $\{\rho^{(0)}_X \}$ (and likewise we also can define $\rho^{(0)}_{A}$). Allowing the active subsystem A to relax by submitting  $\rho_{\mathrm{A}}$ to a self-consistent-field optimization in the frozen environment density, $\rho_{\mathrm{env}}^{(0)}$, leads to a relaxed  density $\rho_{A}^{\mathrm{(1)}}$.
In terms of density-based embedding schemes, this approach directly refers to FDE.\cite{Wesolowski1993} 
It yields a relaxed energy for the active subsystem $E_\mathrm{A}[\rho_{\mathrm{A}}^{(1)}]$.

The relaxation of $\rho_{A}^{(0)}$ to $\rho_{A}^{(1)}$ can be done for all  fragments in a step-wise manner until self-consistency to obtain the relaxed densities $\rho_A^{(2)}$ and $\rho_\mathrm{env}^{(2)}$.  
This is denoted a freeze-and-thaw procedure\cite{Wesolowski1996a}. 
Formally, the \textit{mutual polarization} of the densities in the  ground state of the super system is recovered when performing a sufficient number of freeze-and-thaw cycles.   

\par

In contrast to that, the  PE methods approximate both static electrostatics and polarization solely based on frozen densities in the environment, 
 $\rho_\text{env}^{(0)} = \sum_{X \ne A}\rho^{(0)}_X$. An expression for the total energy comparable to eq.~\eqref{eq:int_total_general} can then be obtained through  Rayleigh--Schrödinger perturbation theory,
\begin{align}
	E_{\mathrm{tot}} \approx E^{(0)} + E^{(1)} + E^{(2)}. 	\label{eq:pertubation-0}
\end{align}
By expressing the interaction between the subsystems as perturbations of the energy, the zeroth-order perturbation can be identified as the isolated subsystem energies. Thus, from eq.~\eqref{eq:int_total_general} we identify $E^{(0)} = E_\mathrm{A}[\rho_{\mathrm{A}}^{(0)}] + E_\mathrm{env}[\rho_{\mathrm{env}}^{(0)}]$ 
 and the interaction energy ($E_{\mathrm{int}}[\rho_{\mathrm{A}}^{(0)},\rho_{\mathrm{env}}^{(0)}]$) must therefore come through the  higher-order energy corrections. Indeed, the first-order correction corresponds to  eq.~\eqref{eq:e_coul} with frozen densities, \ie{},~$E^{\mathrm{C}}_{\mathrm{int}}[\rho^{(0)}_{\mathrm{A}},\rho^{(0)}_{\mathrm{env}}] $. The PE model further approximates  $E^{\mathrm{C}}_{\mathrm{int}}[\rho^{(0)}_{\mathrm{A}},\rho^{(0)}_{\mathrm{env}}] $ through a multipole expansion\cite{Stone1981,Stone2013,Olsen2011}, \ie{},  
\begin{align}
	E^{(1)} 
      = E^{\mathrm{C}}_{\mathrm{int}}[\rho^{(0)}_{\mathrm{A}},\rho^{(0)}_{\mathrm{env}}]  
      \approx E^{\mathrm{mult}}_{\mathrm{int}}[\rho^{(0)}_{\mathrm{A}},\rho^{(0)}_{\mathrm{env}}].  
	\label{eq:pertubation-1}
\end{align}	
The multipole expansion employs individual atoms of the subsystems/fragments as expansion points ($\{\mathbf{R}_s\}$ or in short \textit{sites}, $s$). 
The multipole expansion can thus be written as
\begin{align}
E^{\mathrm{mult}}_{\mathrm{int}}[\rho^{(0)}_{\mathrm{A}},\{\rho^{(0)}_{X}\}]  =&    \sum_{X\neq A}\sum_{s\in X}\left(- \int \rho^{(0)}_{\mathrm{A}} (\mathbf{r}_\text{a}) T^{(0)}_{s\mathrm{a}} \text{d}\mathbf{r}_\text{a}  + \sum_{I\in A}Z_I T^{(0)}_{sI}  \right) q_{\mathrm{s}}[\rho_\mathrm{X}^{(0)}]  \notag \\
	 & -    \sum_{X\neq A}\sum_{s\in X}\left( - \int \rho^{(0)}_{\mathrm{A}} (\mathbf{r}_\text{a}) \mathbf{T}^{(1)}_{s\mathrm{a}}\text{d} \mathbf{r}_\text{a}   + \sum_{I\in A} Z_I \mathbf{T}^{(1)}_{sI} \right) \bm{\mu}_{s}[\rho_X^{(0)}]	+ \cdots .
	\label{eq:non-relaxed-embedding-2}
\end{align}
In the above equation, we have defined the interaction operators 
 $\mathbf{T}^{(k)}_{s\mathrm{a}} = \frac{\partial^{k_x + k_y + k_z} }{\partial x^{k_x}_a \partial y^{k_y}_a \partial z^{k_z}_a} |\mathbf{R}_{\mathrm{a}} - \mathbf{R}_{s}|^{-1}$ 
 that describe interactions at point $a$ due to site $s$. Moreover, the multipole moment operator of $k$'th order at site $s$ is defined as 
$
\mathbf{Q}_{s}^{(k)}[\rho^{(0)}_X] 
=
\langle \Psi^{(0)}_{X}\vert \hat{\mathbf{Q}}^{(k)}_{\mathrm{s}}\vert \Psi^{(0)}_{X}\rangle
$. 
The term for zeroth-order moments represents the charge contribution $\hat{\mathbf{Q}}^{(0)}_{\mathrm{s}}=\hat{\mathbf{q}}_{\mathrm{s}}$, the first-order term denotes the dipole contribution 
$\hat{\mathbf{Q}}^{(1)}_{\mathrm{s}}=\hat{\pmb{\mu}}_{\mathrm{s}}$ and so on. 
In the following, we combine the two sums over fragments $X$ and sites $s$ into one sum over all sites $s$. 

The mutual polarization effects are approximately covered by the second-order correction\cite{Stone2013}
\begin{align}
	E^{(2)} & = E_{\text{A}}^{\text{pol}} + E_{\text{env}}^{\text{pol}} + E^{\text{disp}}\label{eq:pertubation-2} .
\end{align}
We focus on the following only on the polarization part, while neglecting $E^{\mathrm{disp}}$. The environment polarization energy, $ E_\mathrm{env}^{\mathrm{pol}}$, 
 can be described as   
\begin{align}
 	E_\mathrm{env}^{\mathrm{pol}} [\rho^{(0)}_{\mathrm{A}}] = -\frac{1}{2} \pmb{\mathcal{E}}^{T} [\rho^{(0)}_{\mathrm{A}}] \cdot \pmb{\mathcal{\mu}}^{\mathrm{ind}}[\rho^{(0)}_\mathrm{A}] 
 	\label{eq:Epol}, 
\end{align}
and $ E_{\text{A}}^{\text{pol}}$ can in principle be obtained analogously. 
This part is, however, inherently included in the QM model for the active system. 
 The field  $\pmb{\mathcal{E}}$
 is defined as the sum of the fields from electrons in system A, nuclei in system A, 
and the multipoles in the environment
\begin{align}
 \pmb{\mathcal{E}}  [\rho^{(0)}_\mathrm{A}]  = \pmb{\mathcal{E}}^{\mathrm{e}}_{\mathrm{A}}
 [\rho^{(0)}_\mathrm{A}] + \pmb{\mathcal{E}}^{\mathrm{n}}_{\mathrm{A}}  + \pmb{\mathcal{E}}^{\mathrm{mult}}_{\mathrm{env}}.
 \label{eq:field-total}
 \end{align}
The induced dipole moment on site $s$ can then be obtained as
\begin{align}
\pmb{\mu}^{\mathrm{ind}}_{s}[\rho^{(0)}_{\mathrm{A}}] = \bm{\alpha}_s \cdot \left( \pmb{\mathcal{E}}_s[\rho^{(0)}_{\mathrm{A}}] + \sum_{s'\neq s}\mathbf{T}^{(2)}_{ss'}\pmb{\mu}^{\mathrm{ind}}_{s'}  \right),
\label{eq:induced-dipole-1}
\end{align}	
where $\bm{\alpha}_s $ is the (static) point-polarizability localized in site $s$ and $\pmb{\mathcal{E}}_s [\rho^{(0)}_{\mathrm{A}}]$ the field in eq.~\eqref{eq:field-total} on site $s$. Note that the induced dipole on site $s$ depends on the field generated from the induced dipoles on all remaining sites. Thus, a self-consistent optimization is required to obtain the induced dipole moment $\pmb{\mu}^{\mathrm{ind}}_s$. This optimization problem can be written as 
\begin{align}
 \pmb{\mu}^{\mathrm{ind}}_{s}[\rho_{\mathrm{A}}^{(0)}] = \sum_{t}\mathbf{R}_{ts} \pmb{\mathcal{E}}_{\mathrm{A},s}[\rho^{(0)}_\mathrm{A}], 
 \label{eq:induced-dipole-2}
\end{align}
where the so-called classical response matrix, $\mathbf{R}$, is given as
\begin{align}
\mathbf{R} =
\begin{pmatrix}
 \pmb{\alpha}_{1}^{-1} & -\mathbf{T}_{12}^{(2)} &  \dots &   -\mathbf{T}_{1S}^{(2)}  \\ 
 -\mathbf{T}_{21}^{(2)} & \pmb{\alpha}_{2}^{-1} & \dots &  -\mathbf{T}_{2S}^{(2)} \\
\vdots & \vdots & \ddots & \vdots\\
 -\mathbf{T}_{S1}^{(2)} & -\mathbf{T}_{S2}^{(2)} & \dots & \pmb{\alpha}_{S}^{-1} \\
\end{pmatrix}^{-1}.  
\label{eq:R-matrix}
\end{align}

The total energy (in eq.~\ref{eq:int_total_general}) is now defined by combining eqs.~\eqref{eq:pertubation-0}--\eqref{eq:pertubation-2},
\begin{align}
	E^{\mathrm{PE}}_{\mathrm{tot}}[\rho_{\mathrm{A}}, \rho_{\mathrm{env}}^{(0)}] 
           = E_\mathrm{A}^\mathrm{PE}[\rho_{\mathrm{A}},\rho_{\mathrm{env}}^{(0)}] 
           + E_\mathrm{env}^\mathrm{PE}[\rho_{\mathrm{env}}^{(0)},\rho_{\mathrm{A}} ] +  E^{\mathrm{mult}}_{\mathrm{int}}[\rho_{\mathrm{A}},\rho^{(0)}_{\mathrm{env}}], 
	\label{eq:EtotPEb}  
\end{align}
where we skip the superscript for $\rho_{\mathrm{A}}$ to denote that it is subject to change  in the self-consistent-field (SCF) procedure performed during the optimization of the QM system (note that eq.~\eqref{eq:induced-dipole-2} will then have to be solved within each SCF cycle). The environment density $\rho_\mathrm{env}^{(0)}$ remains the isolated density 
 of the environment fragments (represented by a multipole expansion). 
For consistency with the definition in eq.~\eqref{eq:int_total_general}, we have written the total energy in eq.~\eqref{eq:EtotPEb} as a sum of the energies of the active system A and the environment 
  plus an interaction energy, where we have combined the energy of the isolated subsystem and polarization in the term $
E^\mathrm{PE}_\mathrm{A}[\rho_\mathrm{A},\rho_{\mathrm{env}}^{(0)}] = 
E_\mathrm{A}[\rho_\mathrm{A}] + E_\mathrm{A}^{\mathrm{pol}} [\rho^{(0)}_{\mathrm{env}}]$. 
 The term $E^\mathrm{PE}_\mathrm{env}[\rho_\mathrm{env}^{(0)},\rho_{\mathrm{A}}]$ is defined in a similar fashion. 

With this starting point, a PE embedding potential according to eq.~\eqref{eq:vemb} is derived in Ref.~\cite{bookchapter}. The emanating formalism can be summarized as follows 
\begin{align}
v_{\text{A}}^\mathrm{PE} 
=& 
\frac{\delta E_\mathrm{tot}^{\mathrm{PE}}[\rho_{\mathrm{A}},\rho_\text{env}^{(0)}] }{\delta \rho_{\mathrm{A}}}  
- 
\frac{\delta E_\mathrm{A}^\mathrm{PE}[\rho_{\mathrm{A}}]}{\delta \rho_{\mathrm{A}}} = {v}^{\mathrm{mult}} + {v}^{\mathrm{pol}},
\label{eq:V_PEemb-1}  	
\end{align}
where the two operators are defined as 
\begin{align}
{v}^{\mathrm{mult}} & = \sum_{s}
\sum_{k=0}\frac{(-1)^{|k|}}{k!}
{\mathbf{T}}^{(k)}_{s\mathrm{a}} (\mathbf{r}_\mathrm{a}) \mathbf{Q}^{(k)}_{s} \label{eq:vmult} 
\end{align}
and
\begin{align}
{v}^{\mathrm{pol}} & = 
\sum_{s}\Big( \pmb{\mu}^{\mathrm{ind}}_s [\rho_{\mathrm{A}}]\Bigr)^T \ \pmb{\varepsilon}^{\mathrm{e}}_{\mathrm{A},s}(\mathbf{r}_{\mathrm{a}}), 
\label{eq:vpol}
\end{align}
The field  potential
$\pmb{\varepsilon}^{\mathrm{e}}_{\mathrm{A},s}  (\mathbf{r}_\mathrm{a})$ is the component of the electronic part of the electric field operator at the site $s$, defined in real-space coordinates as,
\begin{align}
\pmb{\varepsilon}^{\mathrm{e}}_{\mathrm{A},s}  (\mathbf{r}_\mathrm{a}) 
= 
\frac{\delta \mathcal{E}^\mathrm{e}_{\mathrm{A},s} [\rho_\mathrm{A}(\mathbf{r}_\mathrm{a})]}{\delta \rho_\mathrm{A} (\mathbf{r}_{\mathrm{a}})}.
\label{eq:et}
\end{align}
where $\mathcal{E}^\mathrm{e}_{\mathrm{A},s} [\rho_\mathrm{A}(\mathbf{r}_\mathrm{a})]$ is the electronic component of the field in system A at site $s$ (cf.~eq.~\ref{eq:field-total}). 
Thus, ${v}^{\mathrm{mult}}$ corresponds to a multipole approximation of eq.~\eqref{eq:emb_rho}, where only $\rho^{(0)}_X$ are employed. 
Similarly, ${v}^{\mathrm{pol}}$ approximates the effect of mutual polarization, \ie{}, moving from $\rho^{(0)}_X$ to $\rho^{(2)}_X$ in eq.~\eqref{eq:emb_rho}. The QM contributions from eq.~\eqref{eq:FDEpot} are not included in standard PE. 

Optical spectra are in this work obtained by linear response theory. To incorporate embedding contributions of the introduced models in local response calculations, we employ the commonly used framework of time-dependent DFT (TD-DFT). 
Therefore, we add the embedding potential $v_\text{emb}$ to the Kohn-Sham operator of the vacuum system $\hat{f}_\text{iso}$ 
\begin{align} 
\hat{f}_\text{tot} = 
\hat{f}_\text{iso} + v_\text{emb},
\label{ftot_emb}  
\end{align} 
 with $v_\text{emb}$ being eqs.~\eqref{eq:FDEpot} or \eqref{eq:V_PEemb-1} for FDE or PE, respectively.  
 Replacing $\hat{f}_\text{iso}$ with $\hat{f}_\text{tot}$  
 in the derivation of the response equations leads to a set of modified response equations,  that are,  
\begin{align}
	\left[
	\begin{pmatrix}
		\phantom{-}\mathbf{A} & \mathbf{B} \\
		-\mathbf{B}^* & -\mathbf{A}^* \\
	\end{pmatrix}
	- \omega \begin{pmatrix}
		~\mathbf{1} & \mathbf{0}~ \\
		~\mathbf{0} & \mathbf{1}~\\
	\end{pmatrix} 
	\right] 
	\begin{pmatrix}
		\mathbf{X} \\ \mathbf{Y} 
	\end{pmatrix} = 0  ,
	\label{eq:tddft-1}
\end{align}
with the excitation energies $\omega$ and %, $d_{ai} = X_{ai}$, $ d_{ia} = Y_{ia}$ and 
\begin{align}
	A_{ai,bj} = &  \delta_{ij} \delta_{ab} \left(\varepsilon_a - \varepsilon_i \right) + B_{ai,jb} \label{eq:tddft-2} \\
	B_{ai,bj} = &  \frac{\partial F^\text{iso}_{ai}}{\partial P_{bj}} 
	+
	\frac{\partial \left\langle \phi_a | {v}^\text{emb} | \phi_i \right\rangle}{\partial P_{bj}}.\label{eq:tddft-3}
\end{align}
$F^\text{iso}_{ai}$ denotes a Fock matrix element of the isolated system and $P_{bj}$ is 
 an element of the density matrix. 
The quantities $\varepsilon_a$ and $\varepsilon_i$ are orbital energies, where occupied orbitals are labelled with $i$ or $j$ and the virtual ones with $a$ or $b$. The orbital energies are eigenvalues of the \textit{total} Fock operator $\hat{f}_\text{tot}$ in eq.~\eqref{ftot_emb}. Thus, part of the environmental contribution enters through these energies. 

For FDE, ${v}^\text{emb}$ can be chosen to rely only on $\{\rho^{(0)}_X\}$ densities, which we will denote \fde\ \nopol{}. An equivalent contribution can be defined for PE (which we denote \pe\ \nopol{}) if only $v^{\mathrm{mult}}$ of eq.~\eqref{eq:V_PEemb-1} is included in $v_{\mathrm{emb}}$. Employing the relaxed densities $\{\rho^{(2)}_X\}$ in eq.~\eqref{eq:vemb} corresponds to including the ground-state polarization and we denote this model \fde{}  \gspol{}. The corresponding model in the PE framework (denoted \pe{} \gspol{}) corresponds to employing both $v^{\mathrm{mult}}$ and $v^{\mathrm{pol}}$ of eq.~\eqref{eq:V_PEemb-1} in $v_{\mathrm{emb}}$,  while neglecting the $v_{\mathrm{emb}}$ part of $B_{ai,bj}$ in eqs.~\eqref{eq:tddft-2} and \eqref{eq:tddft-3}.  

The second term of B requires more attention since the physical content between PE and FDE models is rather different \cite{bookchapter}. 
 The term can be identified as 
\begin{align}
\frac{\partial \left\langle \phi_a | {v}^\text{emb} | \phi_i \right\rangle}{\partial P_{bj}} = 
\left\langle
\phi_a(\mathbf{r}_\mathrm{a})
\phi_b(\mathbf{r}'_\mathrm{a})
\left|
\frac{\delta v^\mathrm{emb}_\mathrm{A}(\mathbf{r}_\mathrm{a})}{\delta \rho_\mathrm{A}(\mathbf{r}'_\mathrm{a})}
\right|
\phi_i(\mathbf{r}_\mathrm{a})
\phi_j(\mathbf{r}'_\mathrm{a})
\right\rangle. 
\label{eq:tddft-4}
\end{align}
The functional derivative ($\frac{\delta v^\mathrm{emb}_\mathrm{A}(\mathbf{r}_\mathrm{a})}{\delta \rho_\mathrm{A}(\mathbf{r}'_\mathrm{a})}$) 
 for the corresponding embedding scheme can be derived from eqs. \eqref{eq:FDEpot} and \eqref{eq:V_PEemb-1} for FDE and PE, respectively.
\\
In the case of FDE, we have a static (ground-state) potential and the  Coulomb terms vanish. Thus, only quantum-mechanical terms contribute \cite{Casida2004}, so that
\begin{align}
\frac{\delta v^\mathrm{FDE}_\mathrm{A}(\mathbf{r}_\mathrm{a})  }{\delta \rho_\mathrm{A}(\mathbf{r}'_\mathrm{a})}
 = &
 \frac{\delta^2 E^\mathrm{QM}[\rho_\mathrm{tot}]} {\delta \rho_\mathrm{tot}(\mathbf{r}_\mathrm{a}) \delta \rho_\mathrm{tot}(\mathbf{r}'_\mathrm{a})}
 -
 \frac{\delta^2 E^\mathrm{QM}[\rho_\mathrm{A}]} {\delta \rho_\mathrm{A}(\mathbf{r}_\mathrm{a}) \delta \rho_\mathrm{A}(\mathbf{r}'_\mathrm{a})} .
\label{eq:FDEpotkernel}
\end{align}
The remaining QM embedding contributions to the response kernel, however, are often small, so that
 the major environmental effect in the electronic transitions and oscillator strengths are results of differences in the canonical orbitals and orbital energies.\cite{Jacob2006,Gomes2008} 
The QM contributions to the embedding in the response kernel are, hence, not included in the numerical examples of the present work.  

For PE, the functional derivative is obtained from eq. \eqref{eq:V_PEemb-1} as\cite{bookchapter}
\begin{align}
  \frac{\delta v^\mathrm{PE}_\mathrm{A}(\mathbf{r}_\mathrm{a})}{\delta \rho_\mathrm{A}(\mathbf{r}'_\mathrm{a})}
 =
 -\sum_{t}\sum_{s} \mathbf{T}^{(1)}_{\mathrm{a}t} (\mathbf{r}'_\mathrm{a})\mathbf{R}_{ts} \mathbf{T}^{(1)}_{\mathrm{a}s} (\mathbf{r}_\mathrm{a}).
 \label{eq:tddft-PE-diffpol}
\end{align}
This contribution can be understood as an (approximate) treatment of differential polarization, \ie{}, the difference in the interaction between the ground-state and excited-state densities with the environment densities. We denote PE models with this effect included as \pe\ \dpol{}. 
There is no corresponding term for FDE models, although extensions have been suggested that include differential polarization\cite{Neugebauer2010,Daday2013,Daday2014,Daday2015,Pal2019,Scholz2021,Harshan2022}. 
Most of them are, however, rather computationally demanding\cite{Neugebauer2010} or require embedded excited-state densities\cite{Daday2013}, which are somewhat tedious to obtain for TD-DFT methods\cite{Daday2014}.

While excitation energies are a fundamental part of a UV-vis spectrum, the associated intensities are often highly important for assignments. The intensity is usually estimated based in the oscillator strength which can also be extracted from eq.~\ref{eq:tddft-1}; for transition $n$, the  oscillator strength, $f_n$, can be calculated 
as 
\begin{align} 
f_n = \frac{2}{3} \omega_n \bm{\mu}_n^2,
\end{align} 
where $\omega_n$ is the excitation frequency and $\bm{\mu}_n^2$ is the transition dipole moment . 
The latter can (for the $\alpha$-component) can be obtained from the  converged response vectors in Eq.~\eqref{eq:tddft-1} 
 as\cite{autschbach2002}  
\begin{align}
\mu_n^\alpha = \frac{1}{\sqrt{\omega_n}} \bm{M}^{\alpha} (\bm{X} + \bm{Y}), \label{eq:polarizability-frequency}
\end{align}
where the $\bm{M}^{\alpha}$ is a vector comprised of the $\alpha = x, y$, and $z$  components with the elements 
\begin{align}
M^{\alpha}_{ai} = - \langle \phi_a | r_{\alpha} | \phi_i \rangle.
\end{align}
Since the introduction of $v^\text{emb}$ in the response equations (eqs.~\ref{eq:tddft-1}--\ref{eq:tddft-3}) also affects the 
 eigenvectors, the embedding also influences the calculated oscillator strengths. 
However, the external field employed to excite the solute also generates an induced dipole on environment sites, that effectively modifies $\bm{M}$. 
 This effect is not included in the standard local embedding schemes but 
 can approximately be accounted for by an external effective field (EEF) term, $\langle \phi_a \vert \hat{V}^{\mathrm{loc}}\vert \phi_i \rangle$, with \cite{List2016,Harshan2022} 
\begin{align}
\hat{V}^{\mathrm{loc}} = \sum_{ts} \mathbf{T}^{(1)}_{t\mathrm{a}}\mathbf{R}_{ts}\mathcal{\bm{E}}^{\mathrm{uni}}_s = \sum_{t}\mathbf{T}^{(1)}_{t\mathrm{a}}\bm{\mu}^{\mathrm{ind}}_{\mathrm{ext},t}, 
\label{eq:eef}
\end{align}
where $\bm{\mu}^{\mathrm{ind}}_{\mathrm{ext},t}$ is the induced dipole to a unit field, $\mathcal{\bm{E}}^{\mathrm{uni}}_s$.

We have summarized the contributions considered in the different embedding approaches in  Tab.~\ref{tab:contribution_overview}, and we refer to the labels used in this table in the following sections. 
\renewcommand*{\arraystretch}{0.5}
\begin{table}%[H]
\caption{Overview of the included contributions in the different embedding models considered here.\label{tab:contribution_overview}}
\centering
\begin{tabular}{lccccccm{3.5cm}}
\toprule
%%---------------------------------------------------------------------------------------------------------------------------------------------------------------------------
Class         &  Label    & Static          & G.s. pol    & QM$^{a)}$   & Diff. pol.  & EEF & Comments   \\%[0.5ex] 
%-------------------------------------------------------------------------------------------------------------------------------------------------------- 
\midrule%[-1.5ex]
QM/classical & \pe\ \nopol{}   & \cmark   & \xmark        & \xmark      & \xmark     &  \xmark   & $v^\text{mult}$ (based on $\{\rho^{(0)}_X\}$), see eqs.~\eqref{eq:V_PEemb-1} and \eqref{eq:vmult}. \\[1.5ex]
QM/QM        & \fde\ \nopol{}  & \cmark   & \xmark        & \cmark      &  \xmark    &  \xmark   & $v_\mathrm{A}^\mathrm{FDE} (\mathbf{r}_a)$ in eq.~\eqref{eq:FDEpot} based on $\{\rho^{(0)}_X\}$. \\[1.5ex]
\midrule%[-1.5ex]
QM/classical & \pe\ \gspol{}   & \cmark   & \cmark        & \xmark     & \xmark      &  \xmark   & $v^\text{mult}$ (based on $\{\rho^{(0)}_X\}$) and $v^\text{pol}$, see eqs.~\eqref{eq:V_PEemb-1} --\eqref{eq:vpol}.  \\%[0.5ex]
QM/QM        & \fde\ \gspol{}  & \cmark   & \cmark        & \cmark     & \xmark      &  \xmark  & $v_\mathrm{A}^\mathrm{FDE} (\mathbf{r}_a) $ (eq.~\ref{eq:FDEpot}) based on $\{\rho^{(2)}_X\}$. \\[1.5ex]
\midrule%[-1.5ex]
QM/classical & \pe\ \dpol{}    & \cmark   & \cmark    & \xmark         & \cmark      &  \xmark  & \pe{} \gspol{} and additionally eq.~\eqref{eq:tddft-PE-diffpol}.   \\[2ex]
QM/classical & \pe\ \eef{}          & \cmark   & \cmark    & \xmark         & \cmark      &  \cmark &  \pe\ \dpol{} with modified  dipole transition moments, eq.~\eqref{eq:eef}   \\[1.5ex]
\bottomrule% \hline
%-----------------------------------------------------------------------------------------------------------------------------------------------------------------------------
\end{tabular} 
\begin{minipage}{\linewidth}
\setstretch{0.8}
\centering
\begin{footnotesize}
a) Can be included via orbital-free DFT. Note that we do not consider the response kernel, eq.~\eqref{eq:FDEpotkernel}, in this work.\\
\end{footnotesize}
\end{minipage}
\end{table}

\subsection{Computational Setup\label{sec:implementation}}

The common theoretical  comparison of density-based QM/MM embedding and PE is only a first step. We also aim for a setup that allows a one-to-one  comparison between the two embedding models in practical calculations. Our setup is shown in fig. \ref{fig:implementation}. 
In the following, we describe the employed workflows. \\ 

\tikzstyle{pro} = [rectangle, rounded corners, minimum height=1.3cm , minimum width=0.5cm, 
                   text centered, draw=luhblue, fill=luhblue, 
                   text width=2cm, text=white
                  ]
\tikzstyle{pro2} = [rectangle, rounded corners, minimum height=1.3cm , minimum width=3.0cm, 
                   text centered, draw=luhblue, fill=luhblue, 
                   text width=2.8cm, text=white
                  ]

\tikzstyle{check} = [rectangle, minimum width=2cm, minimum height=1.1cm, text centered, draw=black, fill=luhgreen, text width=2.8cm]
\tikzstyle{bg} = [rectangle, minimum width=8cm, minimum height=3cm, text centered, draw=black, fill=luhgreen]
% orange!30
\tikzstyle{decision} = [diamond, minimum width=3cm, minimum height=1cm, text centered, draw=black, fill=green!30]

\tikzstyle{arrow} = [thick,->,>=stealth]

%%%%%%%%%%%%%%%%%%%%%%%%%%%%%%%%%%%%%%%%%%%%%%%%%%%%%%%%
%% PE
%%%%%%%%%%%%%%%%%%%%%%%%%%%%%%%%%%%%%%%%%%%%%%%%%%%%%%%%
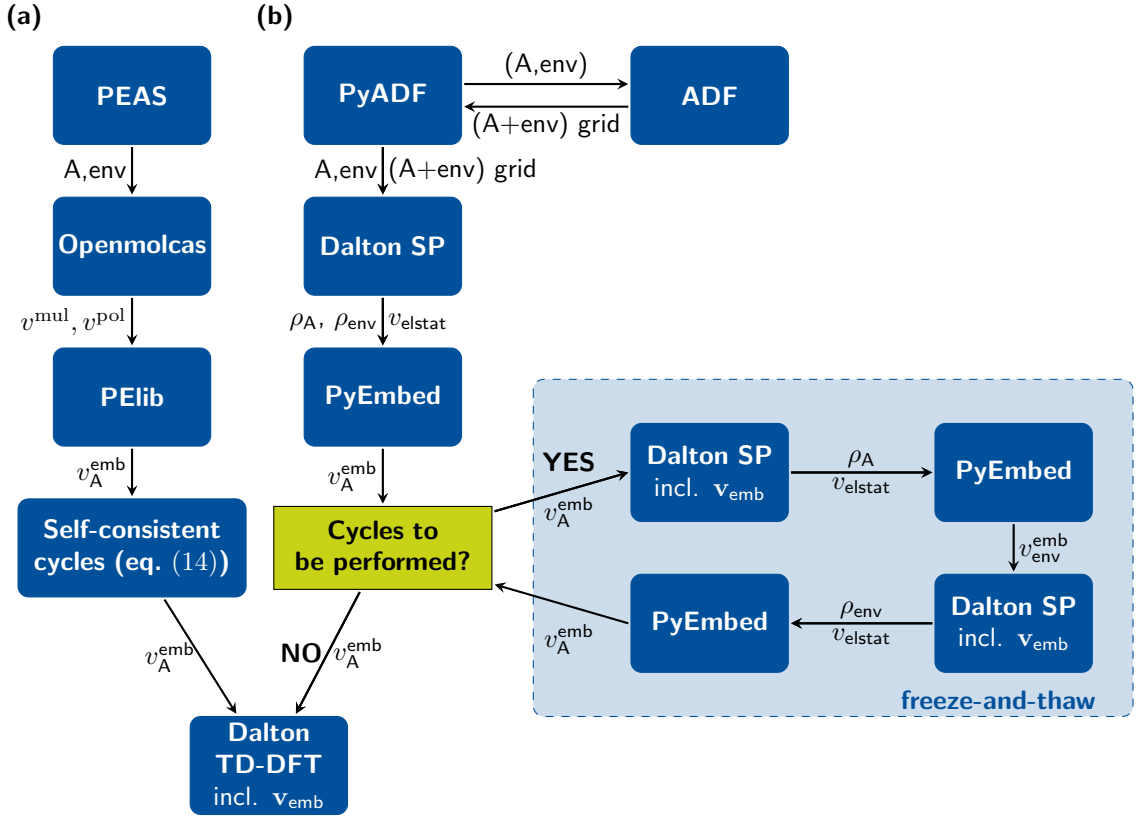
\begin{figure}
\begin{tikzpicture}[node distance=2cm, every node/.style={inner sep=1,outer sep=1}]
\node (peas1) [pro]                 {\textbf{PEAS}};
\path (peas1.north west)+(-0.35,0.35) node (a) {\textbf{(a)}};
% \node (a)     [left of=peas1]                {\textbf{(a)}}
\node (om)    [pro, below of=peas1] {\textbf{Openmolcas}};
\node (peas2) [pro, below of=om]    {\textbf{PElib}};
\node (pe_cycles) [pro2, below of=peas2]    {\textbf{Self-consistent\\ cycles (eq.~\eqref{eq:induced-dipole-1})}};
\node (tddft1) [pro, below of=peas2, below = 2.2cm, xshift=1.8cm]{\textbf{Dalton TD-DFT}\\incl. $\mathbf{v_\text{emb}}$};%xshift=2.3cm, yshift=1cm

\draw [arrow] (peas1) -- node[anchor=east] {A,env}                                (om);
\draw [arrow] (om)    -- node[anchor=east] {$v^{\mathrm{mul}},v^{\mathrm{pol}}$}           (peas2);
\draw [arrow] (peas2) -- node[anchor=east] {$v_{\text{A}}^{\text{emb}}$} (pe_cycles);
\draw [arrow] (pe_cycles) -- node[anchor=east] {$v_{\text{A}}^{\text{emb}}$} (tddft1);
%%%%%%%%%%%%%%%%%%%%%%%%%%%%%%%%%%%%%%%%%%%%%%%%%%%%%%%%
%% FDE
%%%%%%%%%%%%%%%%%%%%%%%%%%%%%%%%%%%%%%%%%%%%%%%%%%%%%%%%
\node (pyadf)    [pro, right of=peas1, xshift=1.3cm]                                 {\textbf{PyADF}};
\path (pyadf.north west)+(-0.35,0.35) node (a) {\textbf{(b)}};
% ADF 
\node (adf)      [pro, right of=pyadf, xshift=2.3cm]     {\textbf{ADF}};
% PYADF WORKFLOW
\node (sp)       [pro, below of=pyadf]                 {\textbf{Dalton SP}};
\node (pyembed)  [pro, below of=sp]                    {\textbf{PyEmbed}};
\node (cycles)   [check, below of=pyembed]             {\textbf{Cycles to \\ be performed?}};
\node (sp2)      [pro,  right of=cycles, xshift=2.3cm, yshift=1cm]    {\textbf{Dalton SP}\\incl. $\mathbf{v_\text{emb}}$};
\node (pyembed2) [pro, right of=sp2, xshift=2cm]       {\textbf{PyEmbed}};
\node (sp3)      [pro, below of=pyembed2]              {\textbf{Dalton SP}\\incl. $\mathbf{v_\text{emb}}$};
\node (pyembed3) [pro, left of=sp3, xshift=-2cm]       {\textbf{PyEmbed}};
\begin{pgfonlayer}{background}
    % Left-top corner of the background rectangle
    \path (sp2.west |- sp2.north)+(-1.25,0.55) node (a1) {};
    % Right-bottom corner of the background rectanle
    \path (sp3.east |- sp3.south)+(+0.5,-0.55) node (a2) {};
    % Draw the background
    \path[fill=luhblue20, rounded corners, draw=luhblue, dashed] (a1) rectangle (a2);
    \path[text=luhblue] (a2.north west)+(-1.7,0.15) node (a) {\textbf{freeze-and-thaw}};
\end{pgfonlayer}

%% ARROWS
\draw [arrow] (pyadf) -- node[anchor=east] {A,env} (sp);
\draw [arrow] (pyadf) -- node[anchor=west] {(A+env) grid} (sp);
% ADF
\draw [arrow] ([shift={(0,0.15)}]pyadf.east) -- node[anchor=south] {(A,env)} ([shift={(0,0.15)}]adf.west);
\draw [arrow] ([shift={(0,-0.15)}]adf.west)  -- node[anchor=north] {(A+env) grid} ([shift={(0,-0.15)}]pyadf.east);
\draw [arrow] (sp)    -- node[anchor=west] {$v_{\text{elstat}}$} (pyembed);
\draw [arrow] (sp)    -- node[anchor=east] {$\rho_\text{A}$, $\rho_\text{env}$} (pyembed);
\draw [arrow] (pyembed) -- node[anchor=east] {$v_{\text{A}}^{\text{emb}}$} (cycles);
\draw [arrow] (cycles) -- node[anchor=east] {\textbf{NO}} (tddft1);
\draw [arrow] (cycles) -- node[anchor=west] {$v_{\text{A}}^{\text{emb}}$} (tddft1);

%%%%% cycles
\draw [arrow] ([shift={(0,-0.1)}]cycles.north east) -- node[anchor=south, yshift=0.2cm, xshift=0.1cm] {\textbf{YES}}(sp2.west);
\draw [arrow] ([shift={(0,-0.1)}]cycles.north east) -- node[anchor=north, yshift=0cm, xshift=0.1cm] {$v_{\text{A}}^{\text{emb}}$}(sp2.west);
% nach oben rechts
\draw [arrow] (sp2) -- node[anchor=south] {$\rho_\text{A}$} (pyembed2);
\draw [arrow] (sp2) -- node[anchor=north] {$v_{\text{elstat}}$} (pyembed2);
% nach unten rechts
\draw [arrow] (pyembed2) -- node[anchor=west] {$v_{\text{env}}^{\text{emb}}$} (sp3);
% nach unten links
\draw [arrow] (sp3) -- node[anchor=south] {$\rho_\text{env}$} (pyembed3);
\draw [arrow] (sp3) -- node[anchor=north] {$v_{\text{elstat}}$} (pyembed3);
% nach cycles
\draw [arrow] (pyembed3.west) -- node[anchor=north east, yshift=-0.2cm, xshift=0.5cm] {$v_{\text{A}}^{\text{emb}}$} ([shift={(0,+0.1)}]cycles.south east);
\end{tikzpicture}

\caption{\label{fig:implementation} Flowchart of the performed workflows for molecule (A) in the environment of the molecule (env). The steps in dark blue boxes stand for subprograms used for a task. The large arrows indicate results that are passed in between the different programs. In either workflow, the supermolecular structure is passed to PEAS or PyADF, respectively, and split up into subsystems. (a) PE workflow utilizing PEAS and PElib\cite{Olsen2020}, which is calling Openmolcas for subsystem calculations. (b) The FDE workflow uses the PyADF scripting framework, that is calling all programs and managing all results mentioned in this workflow. The green box refers to freeze-and-thaw cycles.}
\end{figure}

The general procedure for the PE model involves the construction of the embedding potential. For this purpose we  employ the PE Assistance Script (PEAS).\cite{PEAS}
The script divides the environment into subsystems/fragments and constructs the densities $\{\rho^{(0}_{X}\}$ for the individual fragments, employing DFT calculations with Openmolcas\cite{Openmolcas2019}. From these densities localized multipoles, $\{\mathbf{Q}_{s}[\rho^{(0)}_X]\}$, and static polarizabilities, $\{\bm{\alpha}^{0}_s \}$, 
 can be derived from the LoProp\cite{Gagliardi2004} 
 method, implemented in Openmolcas. 
PEAS collects multipoles and polarizabilities in a potential file that is employed in the calculation of the excitation energies and oscillator strengths with TD-DFT. Optimization of the ground-state as well solving TD-DFT equations [eqs.~\eqref{eq:tddft-1}--\eqref{eq:tddft-3}] are done while including  the PE potential; this  means that the induced dipole moments in eq.~\eqref{eq:tddft-PE-diffpol} are self-consistently optimized along with the SCF/linear response iterations. PElib handles this self-consistent calculation of the induced dipole moments  and adds the resulting PE contributions to the Kohn-Sham or Kohn-Sham-like matrices used in the SCF or response calculations.\cite{Olsen2020}

We employed four different models of increasing accuracy. (i) \pe{} \nopol{} which only contains multipoles up to quadrupoles and no polarizabilities (eq.~\ref{eq:V_PEemb-1}) (ii) \pe{} \gspol{} where full ground-state polarization is included, but differential polarization (eq.~\ref{eq:tddft-PE-diffpol})
 is ignored in the TD-DFT calculation and (iii) \pe{} \dpol{} including both multipoles, ground-state, and differential polarization (eq.~\ref{eq:tddft-PE-diffpol}). (iv) 
A \pe{} \dpol{} with modified dipole transition moments (eq.~\ref{eq:polarizability-frequency}) due to the effect of the external effective field in eq.~\eqref{eq:eef}. This model is equivalent to (iii) for excitation energies but leads to a change in oscillator strengths, and we denote this model \pe{} \eef{}. 

For the FDE calculations, we employed the PyADF scripting framework (fig. \ref{fig:implementation}b).\cite{Jacob2011,jacob2023} 
A supermolecular DFT integration grid was obtained via the ADF program from AMS2020.103 program suite\cite{ams} to ensure that the grids include the full area of all subsystems to preempt grid artifacts that could affect our comparison. 
Subsequently, ground-state calculations of all subsystems were performed individually in Dalton\cite{Dalton}.
The resulting molecular orbitals from these calculations were then translated
 to electron density and electrostatic potentials on the initially generated grid 
 using the DensityEvaluator module of PyADF. 
The non-additive kinetic and the exchange--correlation term (eq.~\ref{eq:FDEpot}) were then evaluated
 by PyEmbed module of PyADF on the same grid. 
Finally, the embedding potential was obtained by adding the environmental electrostatic potential and the environmental non-additive potential for the kinetic and exchange--correlation term.

For the FDE calculations, we employed two potentials: (i) A static embedding potential from isolated environmental densities (skipping the performance of FDE cycles, see fig. \ref{fig:implementation}b: blue underlaid box, \fde{} \nopol{}, eq.~\ref{eq:FDEpot}). (ii) A potential including mutual polarization via freeze-and-thaw cycles of the active subsystem with environmental fragments in the ground state (\fde{} \gspol{}, eq.~\ref{eq:FDEpot}). 
In the freeze-and-thaw procedure, we employ the DensityEvaluator to write updated density and electrostatic potential 
 of every ground-state subsystem calculation in Dalton on the integration grid. 
This results in an updated embedding potential when evaluating the embedding potential with PyEmbed. 
\\
It should be noted, that the current implementation in Dalton includes the non-additive parts of the embedding potential in the SCF process, but not in the response kernel for the TD-DFT calculation.\\
The above-described framework enables us to dissect the embedding contributions and quantify the  different approximations discussed above.  For this, the presented embedding models were to a large degree implemented in Dalton to allow a fair side-by-side comparison and the stepwise inclusion of polarization effects.

%%%%%%%%%%%%%%%%%%%%%%%%%%%%%%%%%%%%%%%%%%%%%%%
\section{Computational Details\label{sec:comp_details}}
%%%%%%%%%%%%%%%%%%%%%%%%%%%%%%%%%%%%%%%%%%%%%%%
The snapshots of \pna{} were taken from an MD simulation, using the AMBER software\cite{Case2005}. 
We parameterized the \pna{} molecule with the General AMBER force field (GAFF)\cite{wang2004} and RESP charges\cite{Bayly1993} calculated with B3LYP\cite{Becke88,LYP,B3LYP} 6-31+G* basis set\cite{Ditchfield1971,Hehre1972,Francl1982} (with PCM\cite{Tomasi2005} using the dielectric constant of water).
The system was set up with tleap of the Amber package and \pna{} was solvated with 3160~water molecules, represented by the OPC model.\cite{Saeed2014}  
We first ran a minimization using 10000~steps of steepest descent, followed by 10000~steps of conjugate gradient minimization. 
We next equilibrated the system by running a 1~ns (in the NPT ensemble), heating the system from 0 to 298~K (at 1 atm. pressure) over the first 20~ps. 
This was followed by a 100~ns production run, using the NPT ensemble (at 298 K), a Langevin thermostat, and a Monte Carlo barostat. Electrostatics were treated with Particle Mesh Ewald\cite{Darden1993}, and non-bonded interactions were cut-off at 12 \AA. The hydrogen bonds were constrained with the SHAKE algorithm.\cite{ryckaert1977,Miyamoto1992}   
For further calculations we arbitrarily selected seven out of the total one hundred obtained snapshots. For these snapshots, we constructed systems where all environment molecules within 3, 4, 5, and 12~\AA{} of \textit{p}NA were included.   \\
The snapshots of pFTAA were taken from an MD simulation, using the GROMACS software\cite{Abraham2015,
Pall2014,Pronk2013,Hess2008,Van2005,Lindahl2001,Berendsen1995,Lindahl2019}. 
We parameterized the pFTAA molecule with an adapted CHARMM force field.\cite{Feller2000, Klauda2005, Sjoqvist2011, Sjoqvist2014}
The pFTAA molecule was solvated with 4028 water molecules, represented by the TIP3P model.\cite{Jorgensen1983} 
We first ran a minimization using 50000 steps of steepest descent. %, followed by 10000 steps of conjugate gradient mimimization.
We next equilibrated the system by running a 10 ns (in the NPT ensemble), heating the system from 0 to 300 K (at 1 atm. pressure) over the first 0.2 ps.
All employed snapshots were taken from a 100~ns production run, using the NVT ensemble (at 300~K), a velocity-rescaling thermostat\cite{Bussi2007}, a Berendsen barostat\cite{Berendsen1984} and electrostatics were treated with Particle Mesh Ewald\cite{Sijbers1984}, and non-bonded interactions were cut-off at 10~\AA. All bonds were constrained with the LINCS algorithm.\cite{Hess1997}
For pFTAA, we only consider solvated models with a 3~{\AA} water environment for a selection of eight independent snapshots. 
We note that for some snapshots there are sodium ions in the 3~{\AA} environment, while for others there are no sodium ions  in close proximity of the dye.
\\
The reference calculations were performed with Dalton 2020\cite{Dalton} in a supermolecular TD-DFT calculation with the CAM-B3LYP\cite{CAMB3LYP} xc~functional.
The workflow for the embedding calculation is shown in fig.~\ref{fig:implementation}. 
The construction of the environment potential in the polarizable embedding approach was performed with LoProp\cite{Gagliardi2004} in Openmolcas\cite{Openmolcas2019} in combination with the Polarizable Embedding Assistant Script (PEAS) \cite{Olsen2012}.
In these fragment calculations, the B3LYP\cite{Becke88,LYP,B3LYP} xc~functional was employed together with ANO-type recontractions of the aug-cc-pVDZ or aug-cc-pVTZ basis set, respectively.\cite{Almlof1987, Dunning,Dunning-aug,Olsen2015}
For sodium ions, the ANO-L basis sets were applied.\cite{Pierloot1995}
The linear response calculation for \pna{} was then carried out with Dalton including the constructed PE potential using PElib\cite{Olsen2020}.
For PE pFTAA calculations with sodium counterions, the QM core region was adapted to account for the lack of repulsion via transferable atomic all-electron pseudopotentials for the sodium ions.\cite{MarefatKhah2020}

In the FDE approach, the supermolecular grid with a \enquote{good} Becke grid quality\cite{ADFBeckeGrid} was obtained with the ADF\cite{ADF2001} code and the TD-DFT calculations with the Dalton code via the PyADF scripting environment.\cite{Jacob2011,jacob2023} 
In line with the calculations for the PE approach, the linear response calculations for \textit{p}NA were performed with the CAM-B3LYP xc~functional whereas for the environment molecules, a B3LYP xc~functional was employed. 
In all FDE calculations, the additive xc~functional BP86\cite{Becke88, Perdew86} and the kinetic energy functional PW91k\cite{PW91,PW91k} were applied
 for the non-additive contributions to the embedding potential. %Dalton nonadd nicht im kernel
Three freeze-and-thaw cycles have been used throughout as this setting had been  found to generally yield sufficient results\cite{Kiewisch2008}. %example https://aip.scitation.org/doi/pdf/10.1063/1.2370947
\\
All calculations for \pna{}  were performed with an aug-cc-pVDZ [see supporting information (SI)] or aug-cc-pVTZ basis set. 
For pFTAA an aug-cc-pVDZ basis set was employed in all calculations.
After calculating the five lowest excitations for the reference as well as for the embedding calculations, they were sorted by the oscillator strength of the transition. The strongest $\pi\rightarrow\pi^*$ transition (ensured \via{} inspection of response vectors and orbitals) was chosen to be compared with other results.%HOMO LUMO?

%\newpage
%%%%%%%%%%%%%%%%%%%%%%%%%%%%%%%%%%%%%%%%%%%%%%%
\section{Results and Discussion\label{sec:results}}
%%%%%%%%%%%%%%%%%%%%%%%%%%%%%%%%%%%%%%%%%%%%%%%
We numerically compare calculated excitation energies and oscillator strengths for the models \nopol{}, \gspol{}, \pe{} \dpol{}, and \pe{} \eef{}   introduced in Section \ref{sec:theory} (see tab.~\ref{tab:contribution_overview} for an overview). 
We generally report on shifts, \ie{}, differences in excitation energy or oscillator strength 
 of a solvation model to the vacuum case with the same structure of the dye. 
We denote these shifts as solvatochromic (\s{}-)shifts and \fshifts{} for excitation energies 
 and oscillator strengths, respectively. 
We compare the shifts from embedding models to reference shifts obtained as the difference of a full quantum-chemical result to the vacuum case ($\Delta$\texttt{REF}). 
We generally denote these shifts with $\Delta$, \ie{}, $\Delta$\nopol{} is the shift obtained with the \nopol{} approximation and 
 $\Delta$\gspol{}, and $\Delta$\dpol{} are defined analogously. 
The individual contributions are then defined with respect to the next lower model, \ie{},  
 $\Delta\Delta$\gspol{} = $\Delta$\gspol{}  $-$ $\Delta$\nopol{},
 $\Delta\Delta$\dpol{} = $\Delta$\dpol{}  $-$ $\Delta$\gspol{}, and 
 $\Delta\Delta$\texttt{EEF} = $\Delta$(\eef{})  $-$ $\Delta$\dpol{}. 
Additionally, we define  
 $\Delta\Delta$(\eef{}) = $\Delta$(\eef{})  $-$ $\Delta$\gspol{}. 

When regarding the proportion of the single contributions ($\Delta$\nopol{}, $\Delta\Delta$\gspol{}, $\Delta\Delta$\dpol{}) to the total shift, we reference to the total supermolecular shift ($\Delta${\texttt{REF}}) for both models, whenever available, and to the total \dpol{} shift ($\Delta$\dpol{}) when a supermolecular reference is unavailable
 (\textit{c.f.} tabs.~\ref{table:pftaa_exen}-\ref{table:pftaa_osc_eef}).

\subsection*{\textit{para}-Nitroaniline}
\begin{figure}%[h]
\centering
\includegraphics[width=0.6\textwidth]{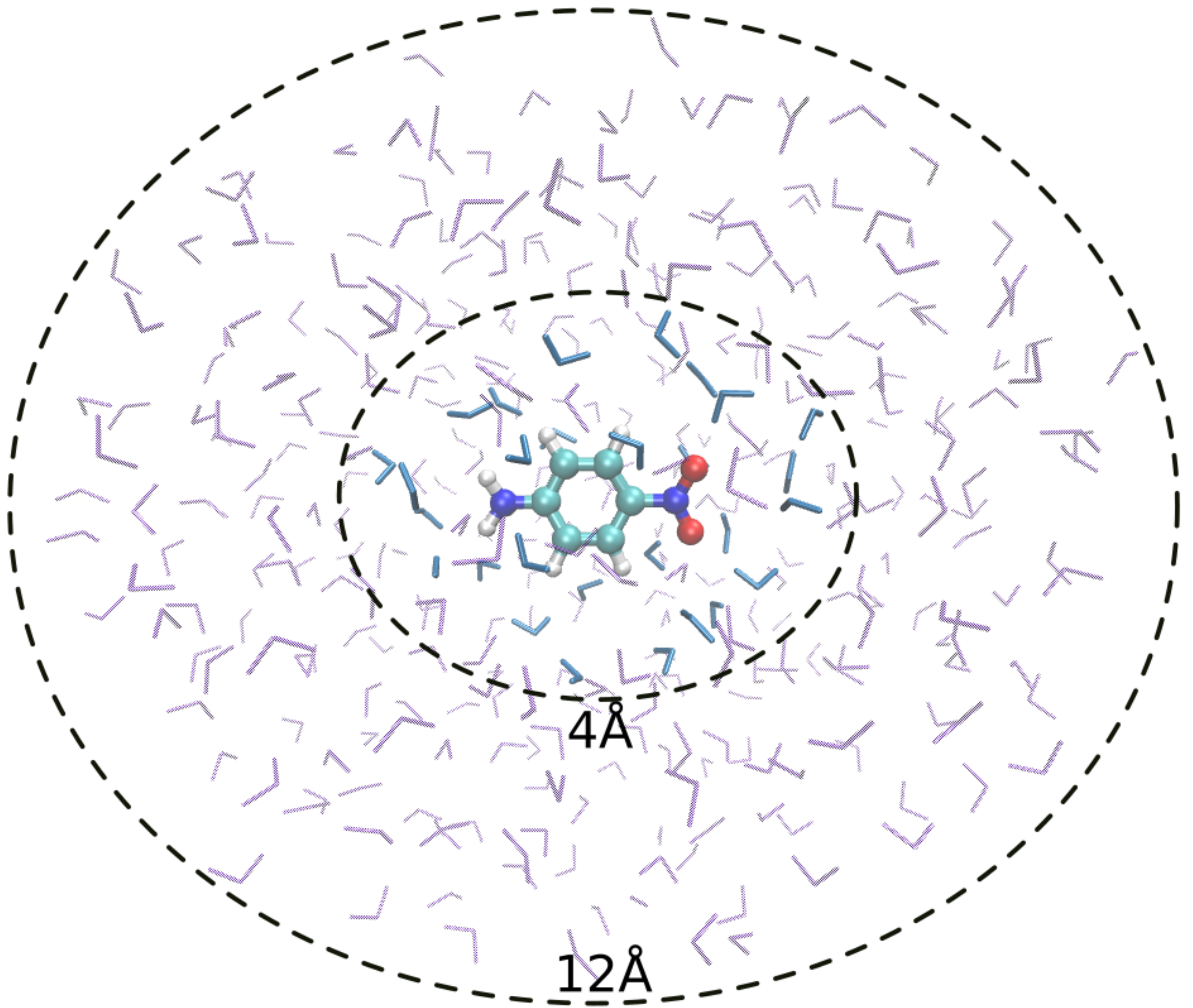}
\caption{\label{fig:pna_md}Example MD configuration of \textit{para}-nitroaniline in a 4~\AA{} and 12~\AA{} water environment selection}
\end{figure}

Our first test system is \textit{para}-nitroaniline (\pna{}) in different water environments (see fig.~\ref{fig:pna_md}).
First, we investigate  environment sizes of 3~\AA{} and 4~\AA{} for seven~snapshots. 
Fig. \ref{fig:pna_solv_en} shows the contributions to the total \s{}-shifts of {\pna} for the different solvent models and compare them to the supermolecular reference. 
It can be seen, that the total \s{}-shift varies largely for the different snapshots, independent of the environment size or embedding scheme used. 
Both, \pe{} and \fde{} models reproduce the changes obtained in the reference calculations qualitatively correctly:  
\begin{figure}%[H]
\centering
\begin{minipage}{0.48\textwidth}
\includegraphics[width=\textwidth]{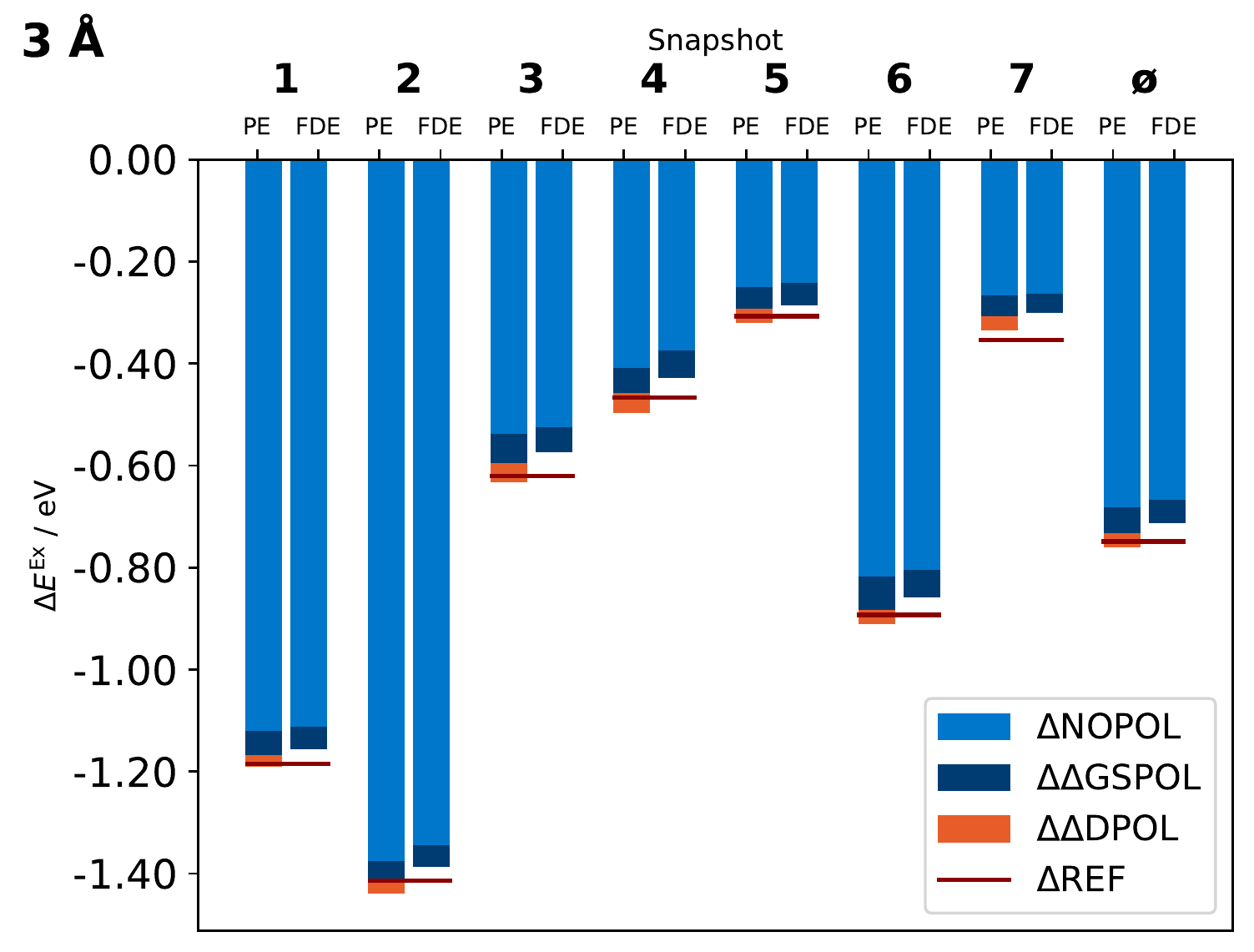}
\end{minipage}\hfill
\begin{minipage}{0.48\textwidth}
\includegraphics[width=\textwidth]{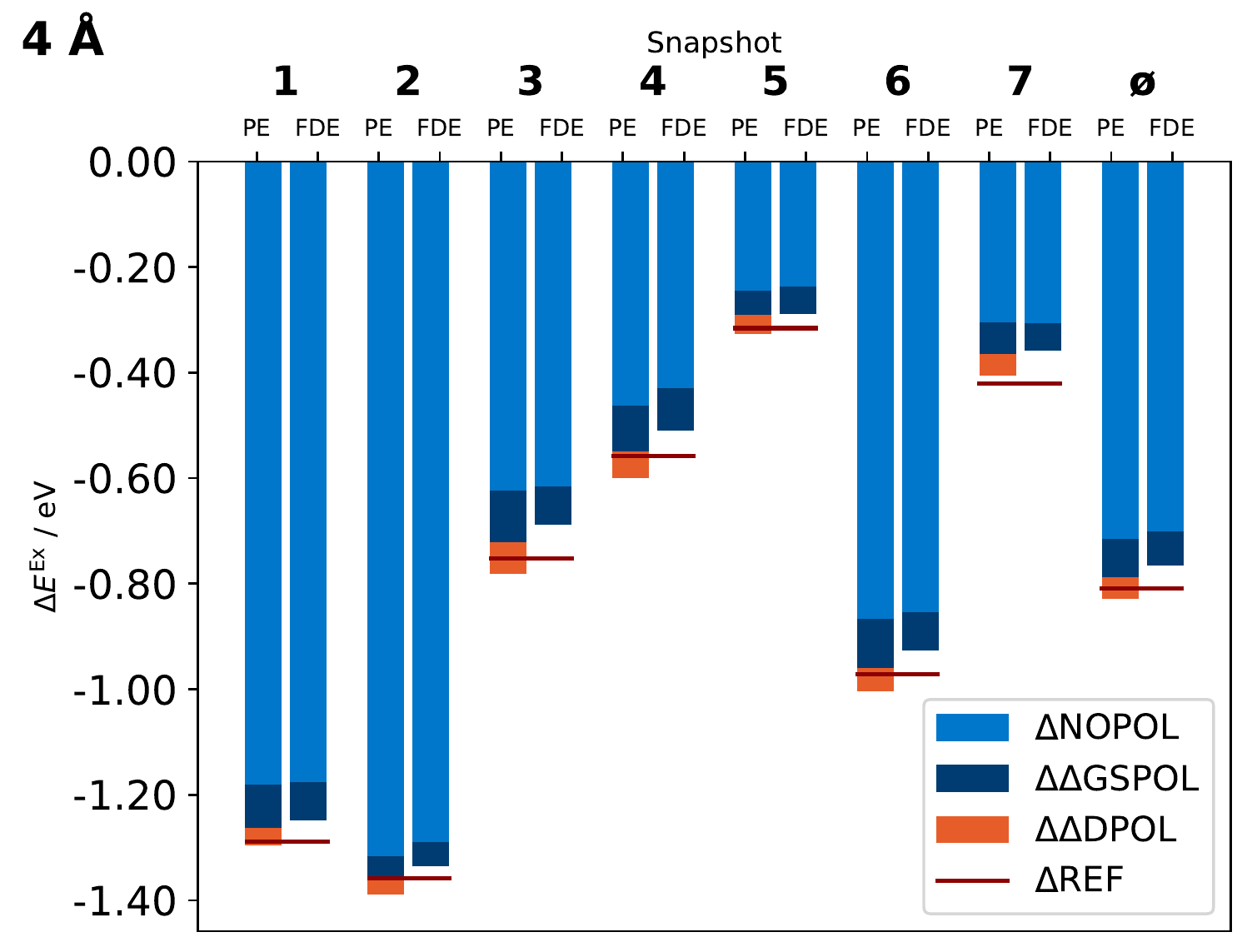}
\end{minipage}

\caption{Contributions from the different models to the total \s{}-shifts and their average for different configurations of \pna{} in 3 and 4~\AA{} environments of water obtained from an MD simulation and subsequently calculated in a PE and FDE framework and different orders of polarization contributions obtained in calculations with a \tz{} basis set.\label{fig:pna_solv_en}}
\end{figure}
For both, \fde{} and \pe{}, $\Delta$\nopol{} is the largest contribution to the \s{}-shift (on average a proportion of  more than 86\%). % and 88\% for \fde{} and \pe{}, respectively). 
Thus, the \gspol{} contribution ($\Delta\Delta$\gspol{}) is small (7\% and 9\% of the supermolecular \s{}-shift for \fde{} and \pe{}, respectively).  
The \pe{} $\Delta\Delta$\dpol{} proportion lies below 5\%. 
Thus, $\Delta$\nopol{} and $\Delta$\gspol{}  for \pe{} as well as \fde{} are in very good agreement with $\Delta$\texttt{REF} (the largest average differences are $-$0.01 eV, see tabs.~S-2.3, S-2.4, S-2.7 and S-2.8 in the SI).   

Ultimately, the total \s-shifts for $\Delta$\gspol{} and $\Delta$\dpol{} are in good agreement with the $\Delta$\texttt{REF} (fig.~\ref{fig:pna_solv_en}).  
$\Delta$\gspol{}  on average slightly underestimates the total \s-shift, whereas adding the $\Delta$\dpol{} leads to an (equally small) overestimation: 
For most snapshots, the $\Delta$\dpol{} from \pe{} is slightly higher than for the $\Delta$\texttt{REF}, with an average  deviation of $-$0.01~eV and $-$0.02~eV for the 3~\AA{} and 4~\AA{} system, respectively 
 (see tables~S-2.3-S-2.4 in the SI for the 3~\AA{} system and tables S-2.7-S-2.8 in the SI for the 4~\AA{} system). 
These differences are smaller than differences we would expect from the differences in the applied xc~functionals (the reference calculation is a full CAM-B3LYP calculation and in the determination of the \pe{} embedding potential B3LYP was employed for environment fragments).

It should be noted that for individual snapshots, the $\Delta\Delta$\gspol{} proportion  can exceed the average considerably. 
This is most pronounced for snapshot~5, which shows an overall small total \s{}-shift: 
Here, the $\Delta\Delta$\gspol{} proportion 
 for both \fde{} and \pe{} constitutes between 13\% and  16\% of the supermolecular \s{}-shift for 3 and 4~\AA{} environments, respectively. 
\pe{} $\Delta\Delta$\dpol{} takes a proportion of 10\% and 13\% in the 3 and 4~\AA{} environments.

We further observe a slight change in the proportions when increasing the environment size: 
When going from 3~{\AA} to 4~{\AA} environment, the average $\Delta\Delta$\gspol{} proportion remains around  7\% of the supermolecular shift for \fde{} and slightly increases from 7\% to 9\% for \pe{}. 
The \pe{} $\Delta\Delta$\dpol{} proportion on average increases from 4\% to 5\%.  
In absolute values, however, these contributions for all environment sizes are rather low, \ie{} at most $-$0.10~eV for $\Delta\Delta$\gspol{} for both \pe{} and \fde{} and $-$0.06~eV for $\Delta\Delta$\dpol{} in \pe{}.

\begin{figure}%[H]
\centering
\includegraphics[width=0.7\textwidth]{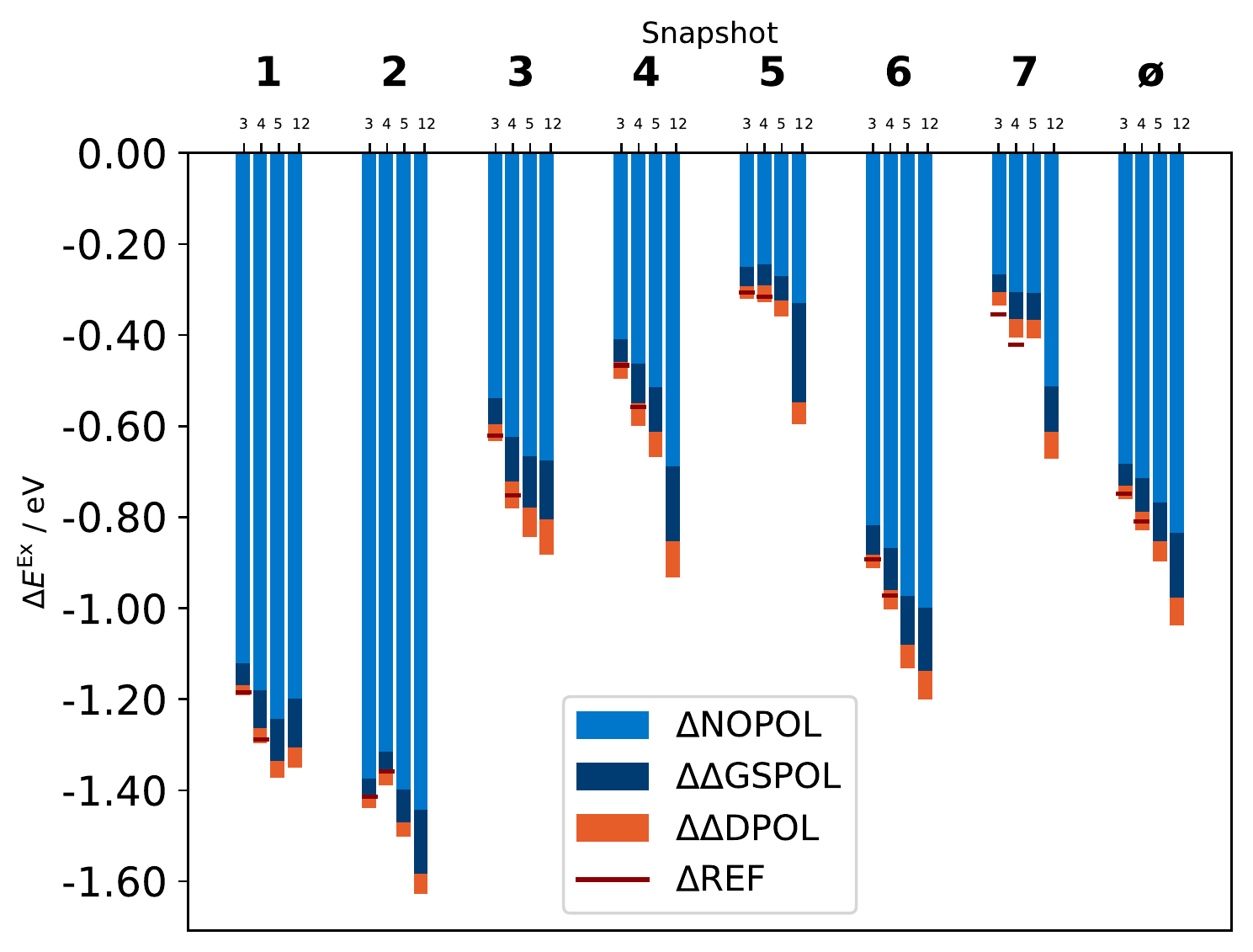}
\caption{Contributions from the different models to the total \s{}-shifts and their average for different configurations of \pna{} in 3, 4, 5, and 12~\AA{} environments of water obtained from an MD simulation and subsequently calculated in a PE framework and different orders of polarization contributions obtained in calculations with a \tz{} basis set. \label{fig:pna_pe_en}}
\end{figure}
For \pe{}, we extend the environment further to 5~\AA{} and a 12~\AA{}. 
The results for the \s{}-shifts from all these calculations are depicted in fig.~\ref{fig:pna_pe_en}. %\\
The overall trend is a distinct increase in the size of the total \s{}-shifts when extending from a 3~\AA{} to a 12~\AA{} environment (\ie{}~the shift becomes more negative): on average it increases by $-$0.27~eV. 
Again, $\Delta$\nopol{} is the largest contribution and it increases  with  enlarged environment size: With the extension from the 3~\AA{} to 4~\AA{} environment it increases by $-$0.02~eV on average, from the 4~\AA{} to the 5~\AA{} environment it increases by $-$0.06~eV on average, and by $-$0.07~eV when further extending to the 12~\AA{} environment. 
$\Delta\Delta$\gspol{} also increases: the increase is  on average $-$0.02~eV  from the 3~\AA{} to 4~\AA{} 
 environment, additional $-$0.02 eV from the 4~\AA{} to the 5~\AA{} environment, and $-$0.05~eV when extending to the 12~\AA{} environment. 
$\Delta\Delta$\dpol{} also shows an increase when going from a 3~\AA{} to a 12~\AA{} environment. 
The absolute contribution on average increases from $-$0.03~eV to $-$0.06~eV. 
However, the average proportion of the total shift does not steadily increase: From a 3 to a 4~\AA{} environment it changes from  $-$0.03~eV (4\% of $\Delta$\dpol{}) to  $-$0.04~eV (5\% of $\Delta$\dpol{}), for a 5~\AA{} environment it decreases to $-$0.04~eV (4\%  of $\Delta$\dpol{}) and increases for a 12~\AA{} environment to $-$0.06~eV (6\% of $\Delta$\dpol{}).
\\
As discussed above, for the snapshots with smaller total shifts, $\Delta\Delta$\gspol{} can exceed the average considerably: 
Here, we again look at snapshot 5 for which the $\Delta\Delta$\gspol{} proportion changes from  13\% %of $\Delta$\ \erik{$\Delta\dpol$?} 
 ($-$0.05~eV ) in a 3~\AA{} environment to 15\% %of $\Delta$\dpol{} 
 ($-$0.05~eV) in a 4~\AA{} environment, 14\%  %of $\Delta$\dpol{} 
 ($-$0.05~eV) in a 5~\AA{} environment, and 37\%  %of $\Delta$\dpol{} 
 ($-$0.22~eV)  in a 12~\AA{} environment, 
where all percentages refer to the $\Delta$ \dpol{} shift. 
Thus, for this particular snapshot, the $\Delta$\nopol{} accounts for 55\% ($-$0.33~eV) of $\Delta$\dpol{} for a 12~\AA{} environment model.

According to previous studies on \pna{} in a water environment with an EOM-CCSD/EFP scheme, Slipchenko \textit{et al.}\cite{Slipchenko2010} found the $\Delta$\nopol{} proportion of the excitation energy to be of similar amount (80\%) as was obtained in our calculations (86--88\%).
The $\Delta\Delta$\dpol{} proportion was determined to be 3--8\% which is in good agreement with our result of 4--5\%.  In their study, increasing the number of water molecules used in the solvation (2--6 molecules) led to an increase in $\Delta\Delta$\dpol{}, similar to the increase observed in our results. % (0.9\%).% THIS IS REFERENCED TO SUPERMOLECULAR
\\
In a study by Sneskov \textit{et al.}\cite{Sneskov2011},  the average polarization contribution was obtained from 100~snapshots. Both the $\Delta\Delta$\gspol{} and the $\Delta\Delta$\dpol{} proportion are  higher than in the present study's result: 19-21\% and 13\% for the $\Delta\Delta$\gspol{} and $\Delta\Delta$\dpol{} proportion, respectively. These authors also noted that the variation of these values is partially dependent on the individual snapshot. %proportion referenced to our highest model: here dpol
In and FDE context, absolute values for $\Delta\Delta$\dpol{} were obtained from mutual optimization with excited-state densities a the study of Daday \textit{et al.}\cite{Daday2013}. The magnitudes  (0.01--0.22 or $-$0.02--0.15 eV, depending on the description of the excited-state density) 
 are similar to our results for $\Delta\Delta$\dpol{} ranging between 0.02--0.06~eV.

\begin{figure}%[H]
\centering
\begin{minipage}{0.48\textwidth}
\includegraphics[width=\textwidth]{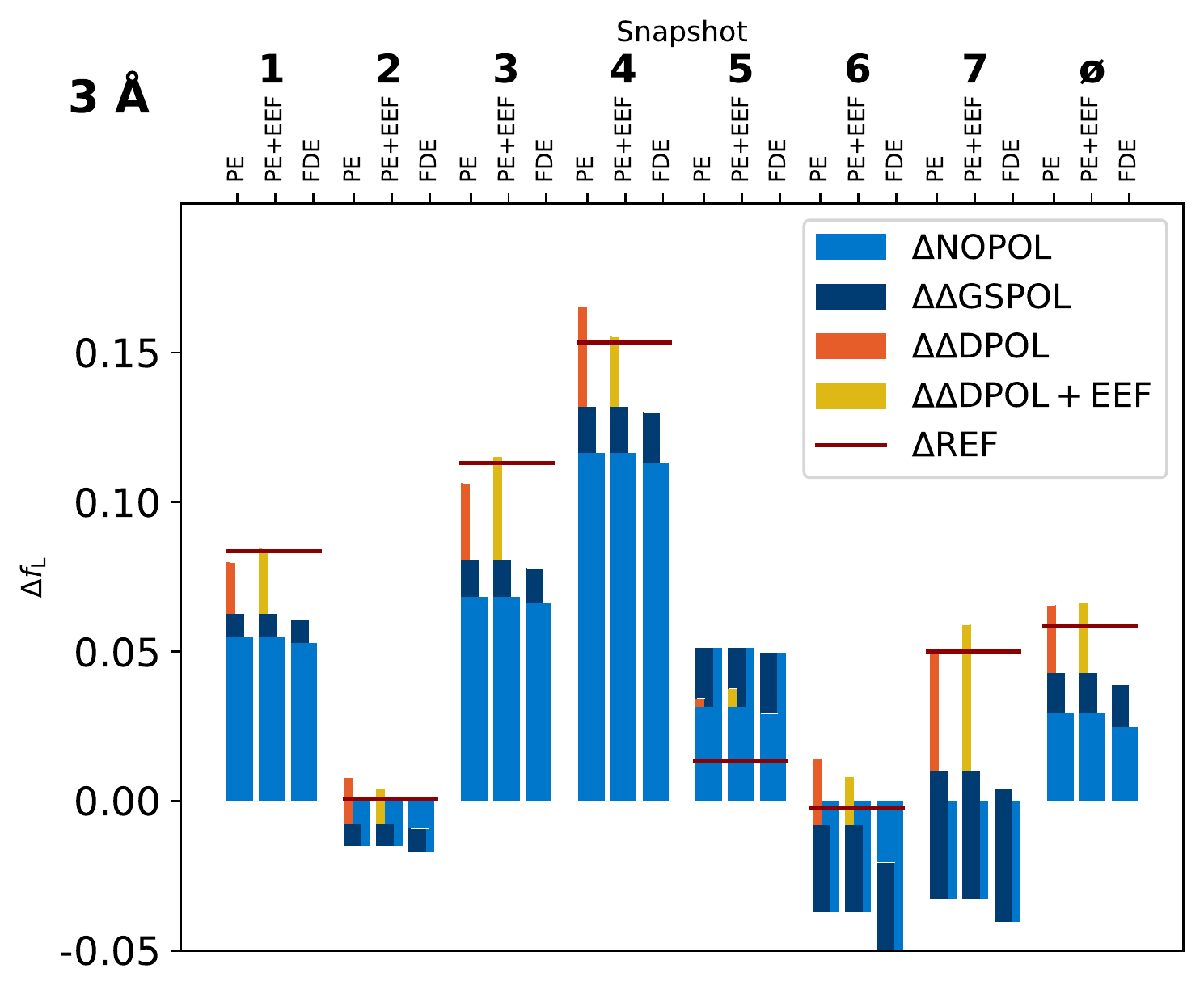}
\end{minipage}\hfill
\begin{minipage}{0.48\textwidth}
\includegraphics[width=\textwidth]{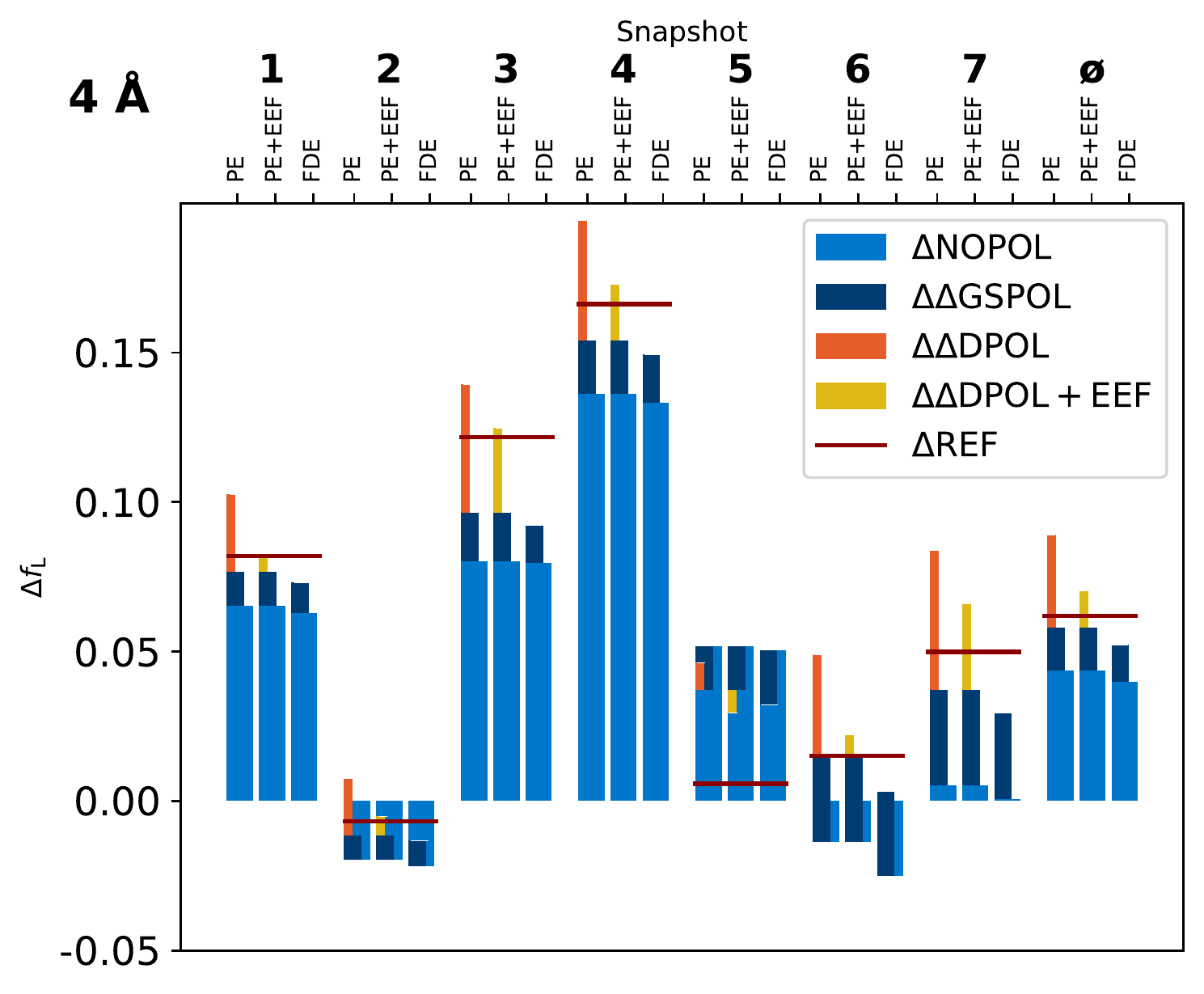}
\end{minipage}

\caption{Contributions from the different models to the total \fshifts{} and their average for different configurations of \pna{} in 3 and 4~\AA{} environments of water obtained from an MD simulation and subsequently calculated in a PE and FDE framework, different orders of polarization contributions and added EEF effects obtained in calculations with a \tz{} basis set. (Full data in tabs. S-2.15, S-2.16, S-2.19, S-2.20)\label{fig:pna_solv_osc_eef}}
\end{figure}
Equivalently to the \s-shifts discussed above, fig.~\ref{fig:pna_solv_osc_eef} displays the change in the oscillator strength ($\mathcal{F}$-shift) of the strongest ($\pi \rightarrow \pi^*$) transition for the various \pe{} and \fde{} solvation models.  We find that the $\mathcal{F}$-shift of the reference calculations displays larger sensitivity than the \s{}-shifts with respect to both, the size of the system and snapshot. This is in line with previous comparisons of different electronic structure methods, showing oscillator strengths to be more sensitive to the employed electronic structure methods\cite{Hedegard2017a}.  

In contrast to the discussion of \s{}-shifts, we here omit the presentation of single contributions ($\Delta$\nopol{}, $\Delta\Delta$\gspol{}, and $\Delta\Delta$\dpol{}) as percentage of the total shift since the oscillator strengths are generally smaller than the excitations energies and even small changes can lead to large percent-wise changes.  
\\
The reference  \fshifts{} are on average 0.06 for both 3 and 4 \AA. The \fde{} and \pe{} \nopol{} models both give an average  \fshift{} of 0.03 for 3 {\AA} and 0.04 for 4 {\AA}. 
Generally, $\Delta$\nopol{} is estimated similarly by \fde{} and \pe{}, the largest deviation being 0.01. 
$\Delta$\nopol{} is \textit{often} the largest contribution to the total \fshift{}, but is much less dominant compared to the $\mathcal{S}$-shift. Notably, the     
$\Delta$\nopol{} results alone are often rather far from the total shifts (most obvious in snapshots 6 and 7 for both 3 and 4 \AA). 
$\Delta\Delta$\gspol{} generally improves the results for \pe{} and \fde{} similarly: 
The largest deviation between \fde{} and \pe{} amounts to less than 0.01 for both the 3~\AA{} and 4~\AA{} systems.

In contrast to the \s{}-shift, $\Delta\Delta$\dpol{} can be rather large for \fshift{}s: 
In some cases (see snapshots 2
 and 6) both $\Delta\Delta$\gspol{} (for \fde{} and \pe{}) and $\Delta\Delta$\dpol{} (for \pe{})  correct the {\fshift} in the opposite direction of $\Delta$\nopol{},  
 where $\Delta$\gspol{} is in better agreement with $\Delta$\texttt{REF} than $\Delta$\dpol{}. This occurs both for snapshots 2 and 6 and on average. 
Especially for the 4~{\AA} environment, we observe large $\Delta$\dpol{} values.  
This over-correction of $\Delta$\dpol{} led us to investigate local field effects on the oscillator strength by means of effective external field (fig.~\ref{fig:pna_solv_osc_eef}).
While for the 3 {\AA} system on average only a small  increase in {\fshift} can be observed (below 0.01), the total {\fshift} for the 4~\AA{} system decreases significantly (for all snapshots), leading to an improved result compared to the reference:  The average deviation is 0.03 for $\Delta$\dpol{} compared to and less than 0.01 for $\Delta$\eef{}.

We finally note that all the discussed results are obtained with \tz{} but for calculations with an \dz{} basis set, the same trends can be observed 
 (see figs.~SI-2.2-SI-2.3 and tabs.~SI-2.1-SI-2.23 in the SI).
The proportions of the single contributions are in line with those obtained with an \tz{} basis set.

In summary,  we observe similar results for \fde{} \gspol{} and \pe{} \gspol{}, suggesting that the additional quantum-mechanical contribution and real-space treatment in FDE have only a minor effect in this case.
The supermolecular reference of excitation energies of \pna{} in the 3~\AA{} and 4~\AA{} water environment  is well in line with FDE results as well as the PE values with or without including differential polarization effects.
When going to larger systems sizes within the \pe{} model, we observe an increased $\Delta\Delta$\gspol{} proportion to the \s{}-shift, but an unclear trend for the differential polarization. 
\\
The observations for the oscillator strengths are similar, though not identical to those for the electronic excitation energies.
In particular, we observe a larger snapshot dependence and the differential polarization contribution in the PE calculations is larger than for the excitation energies. 
We further observe an overcorrection for the 4~{\AA} environments due $\Delta\Delta$\dpol{}, which can be largely cancelled  by accounting for local field effects (EEF).

%\pagebreak
\subsubsection*{Pentameric formyl thiophene acetic acid (pFTAA)}

\begin{figure}
\includegraphics[width=0.45\textwidth]{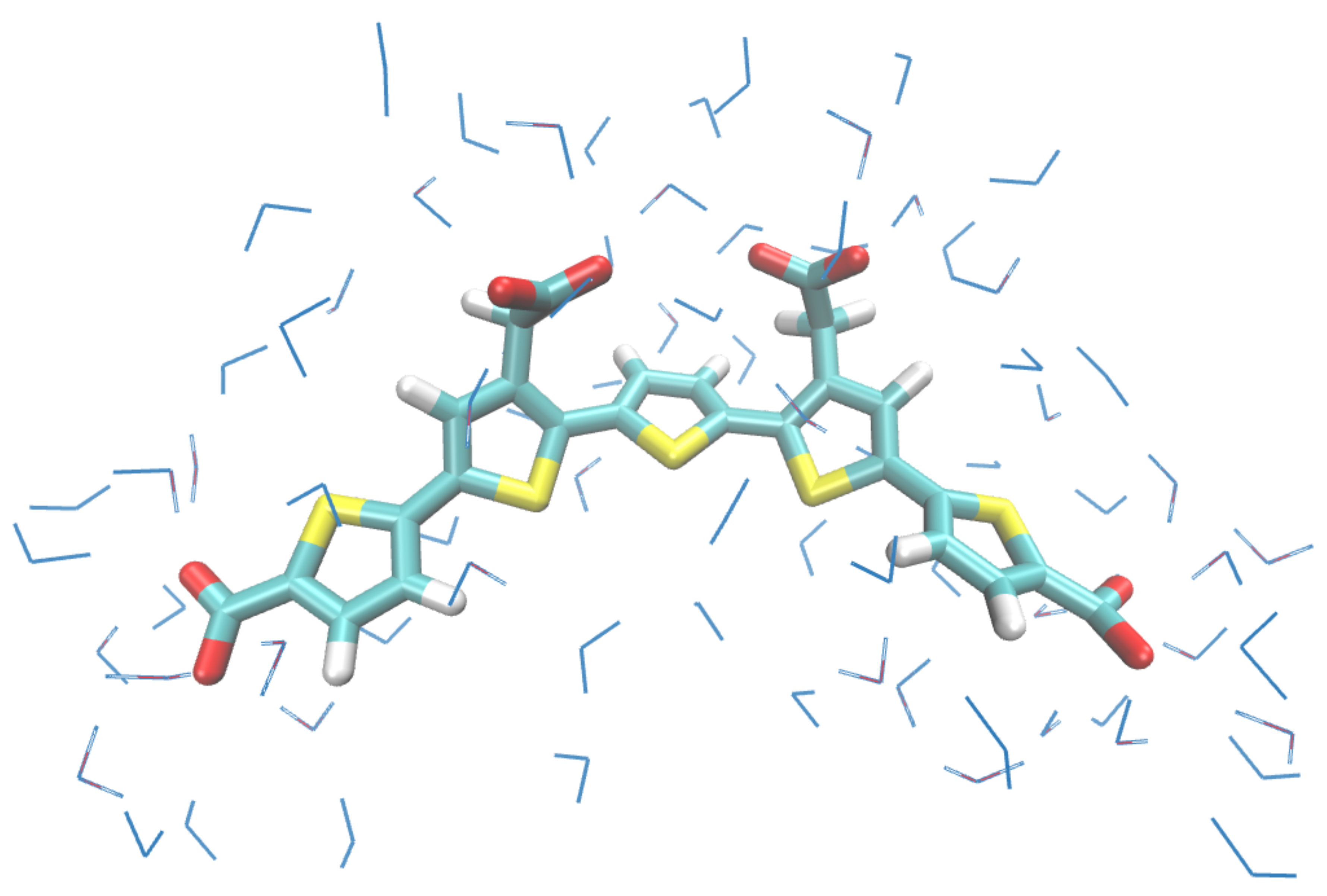}
\hspace{0.1\textwidth}
\includegraphics[width=0.45\textwidth]{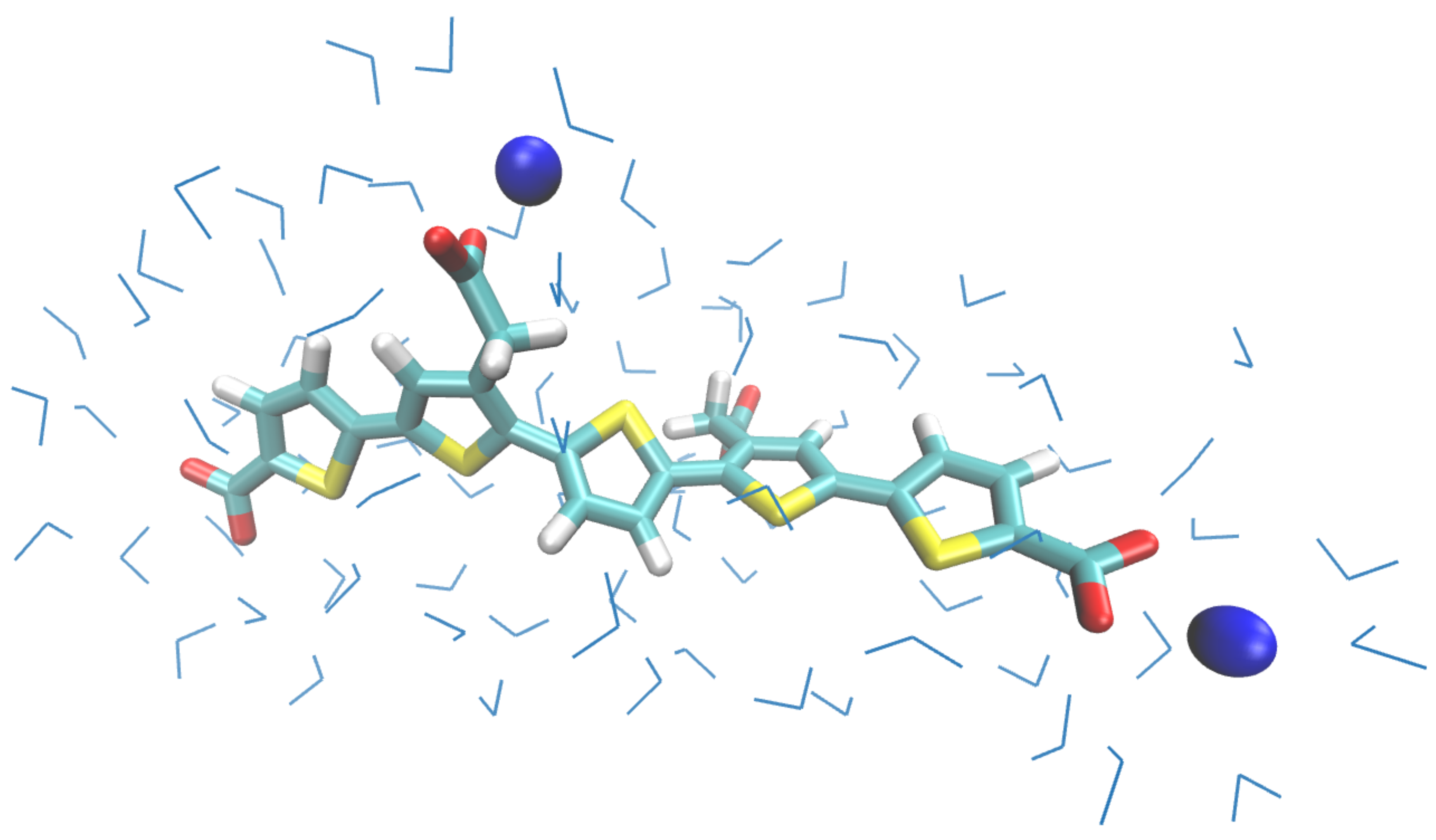}
\caption{\label{fig:pftaa_md}Selected snapshots of pFTAA in a pure 3~\AA{} water environment (left, snapshot 1) and additionally including sodium ions in close vicinity to pFTAA (right, snapshot 2).\label{fig:pftaa_snap}}
\end{figure}
\renewcommand{\arraystretch}{0.7} % default 1 (factor)
\renewcommand{\arraystretch}{2} % default 1 (factor)
\setlength{\tabcolsep}{3pt} % default 6pt
\setlength{\tabcolsep}{9pt} % default 6pt

\begin{table}
	\caption{Contributions from the different models to the total \s{}-shift $\mathcal{S}$ in eV for different configurations of pFTAA in a 3~\AA{} environment of water obtained from an MD simulation and subsequently calculated in a PE and FDE framework and different orders of polarization contributions. Snapshots marked with $^*$ incorporate pseudopotentials in the PE calculations.}
	\begin{center}
	\resizebox{\textwidth}{!}{
		\begin{tabular}{c | c  c   c  c c c c  c c c  c}\toprule
			\multirow[c]{2}{*}{\backslashbox{snap}{$\mathcal{S}$}}    & \multicolumn{3}{c}{\textbf{FDE}}&& \multicolumn{5}{c}{\textbf{PE}} && 
			\multirow{2}{*}{$\Delta$\textbf{REF}} \\
\cline{2-4}\cline{6-10}

			                             & $\mathrm{\Delta NOPOL}$ & $\mathrm{\Delta\Delta GSPOL}$ & $\mathrm{\Delta GSPOL}$ && $\mathrm{\Delta NOPOL}$ & $\mathrm{\Delta\Delta GSPOL}$ & $\mathrm{\Delta\Delta DPOL}$ & $\mathrm{\Delta GSPOL}$  & $\mathrm{\Delta DPOL}$ && 			                             \\
			\hline
%			Snap 40
			1 & 0.15  & 0.01 & 0.17    &&  0.16 & 0.01 & -0.03 & 0.17 & 0.14 && 0.14 \\
%			Snap 20
			2$^*$ & 0.04  & 0.02 & 0.07    &&  0.01 & 0.05 & -0.08 & 0.07 & 0.04 && 0.03 \\
			\bottomrule
		\end{tabular}
		}
	\end{center}
	\label{table:pftaa_exen}
\end{table}

Our second test case, pFTAA, is in contrast to \pna{} highly negatively charged (4$-$).
It is hence, more challenged by possible electron-spill-out effects, which makes it difficult to describe classical models like \pe{}.  
The \fde{} is expected to be less prone to electron-spill-out effects due to the approximate quantum contributions in the embedding potential 
(\textit{c.f. } eq.~\eqref{eq:FDEpot}).\cite{Stefanovich1996, Laio2002} 
Indeed, we found that in the five snapshots that contained sodium cations close to the pFTAA chromophore, the standard \pe{} model broke down. The electron spill-out was revealed by analyzing the contributing orbitals in the response solution vectors.
We counteracted the electron spill-out in these cases  by placing atomic pseudopotentials on sodium ions.\cite{MarefatKhah2020}
This led in all cases to meaningful results. 
The \pe{} results for all these snapshots can be found in tabs. SI-3.2 and SI-3.4 in the SI.

Here, we focus on the discussion of two, representative snapshots: one with and one without a sodium ion in close proximity and again compare the performance of the \pe{} and \fde{} models (see tab~\ref{table:pftaa_exen}). 
Again, the supersystem reference shifts ($\Delta$\texttt{REF})  are well reproduced for both the \fde{} and \pe{} models. 
$\Delta$\nopol{} is by far the largest contribution, while the $\Delta\Delta$\gspol{} is small. Both, $\Delta$\nopol{} and $\Delta\Delta$\gspol{}, 
  are close to identical for \fde{} and \pe{}. 
$\Delta\Delta$\dpol{} is of similar magnitude as $\Delta\Delta$\gspol{}  but points in the opposite direction.

\begin{table}
\caption{Contributions from the different models to the total \fshifts{} for different configurations of pFTAA in a 3~\AA{} environment of water obtained from an MD simulation and subsequently calculated in a PE and FDE framework and different orders of polarization contributions obtained with a \dz{} basis set. Snapshot marked with $^*$ incorporate pseudopotentials in the PE calculations}
	\begin{center}
	\resizebox{\textwidth}{!}{
		\begin{tabular}{c | c  c   c  c c c c c c  c c c  c}\toprule
			\multirow[c]{2}{*}{\backslashbox{snap}{$\mathcal{F}$}}    & \multicolumn{3}{c}{\textbf{FDE}}&& \multicolumn{7}{c}{\textbf{PE}} && \multirow{2}{*}{$\Delta$\textbf{REF}}  \\
\cline{2-4}\cline{6-12}
			                             & $\mathrm{\Delta NOPOL}$ & $\mathrm{\Delta\Delta GSPOL}$ & $\mathrm{\Delta TOT}_\mathrm{GSPOL}$ && $\mathrm{\Delta NOPOL}$ & $\mathrm{\Delta\Delta GSPOL}$ & $\mathrm{\Delta\Delta DPOL}$ & $\mathrm{\Delta\Delta EEF}$ & $\mathrm{\Delta GSPOL}$  & $\mathrm{\Delta DPOL}$ & $\mathrm{\Delta DPOL+EEF}$ &&  \\
			\hline
%            Snap 40 
            1     & 0.294   & 0.053 & 0.348 && 0.320 & 0.060 & 0.063 & -0.029 &  0.380 & 0.443&     0.414     && 0.388 \\
			%            Snap 20
            2$^*$ & 0.140   & 0.044 & 0.183 && 0.143 & 0.024 & 0.052 & -0.076 &  0.167 & 0.219&     0.143     && -0.007 \\
			\bottomrule
		\end{tabular}
		}
	\end{center}
	\label{table:pftaa_osc_eef}
\end{table}

Tab.~\ref{table:pftaa_osc_eef} shows the \fshifts{} for the \pe{} and \fde{} embedding model employing the two snapshots. 
$\Delta$\nopol{} and $\Delta\Delta$\gspol{} are similar for \fde{} and \pe{},  
 where the $\Delta\Delta$\gspol{} contributions are significantly smaller than the $\Delta$\nopol{} contributions. 
$\Delta\Delta$\dpol{} in the \pe{} model for the two investigated snapshots is also rather small and of similar magnitude as the $\Delta\Delta$\gspol{} contributions. 
The $\Delta\Delta$\texttt{EEF} contribution is of similar magnitude as the $\Delta \Delta$\dpol{} contribution but of opposing sign.
\\
For snapshot 1, both $\Delta$\gspol{} and $\Delta$\dpol{} are in reasonable  agreement with the reference value of 0.388: 
For $\Delta$\gspol{} we obtain  deviations of 0.04  and 0.01, respectively for the \fde{} and the \pe{} models. 
As also seen for \pna{}, $\Delta\Delta$\dpol{} overcorrects the \fshift{} (leading to a deviation of 0.05 to the reference), whereas introducing EEF effects ($\Delta$\eef{}) again brings the value closer to the reference. 

%%%%%%%%%%%%%%%%%%%%%%%%%%%%%%%%%%%%%%%%%%%%%%%%
\section{Summary and Conclusions\label{sec:conclusion}}
%%%%%%%%%%%%%%%%%%%%%%%%%%%%%%%%%%%%%%%%%%%%%%%%

We have investigated the influence of different approximations in 
 classical and quantum-based embedding schemes on the excitation energies and oscillator strengths for \textit{p}NA and pFTAA in different sizes of water environments. 
Frozen-density embedding (\fde{}) and polarizable embedding (\pe{}) schemes have been compared. 
To enable a one-to-one comparison of these two methods, we employed an \fde{} framework that complies 
 to a large degree with the \pe{} implementation in Dalton\cite{Olsen2011, Olsen2012,Gao2011}.
In particular, we performed the mutual polarization of subsystems within \fde{} in the 
 PyADF scripting environment\cite{Jacob2011,jacob2023} using Dalton\cite{Dalton} for all (TD-)DFT calculations.
 
With this computational setup at hand, we performed a detailed analysis of the different contributions, \ie{}, static electrostatics (no polarization), ground-state polarization, differential polarization, and quantum-mechanical effects in the \fde{} and \pe{} model to the solvent shift of \pna{} and pFTAA in an explicit water solvent. 
We compared the obtained excitation energies and oscillator strengths to supermolecular TDDFT calculations.

We find that FDE and PE perform similarly with  the inclusion of static environmental densities and ground-state polarization, respectively. 
Since these two contributions are dominating the solvochromatic (\s{}-)shift for both \pna{} and pFTAA, FDE and PE both achieve good agreement with the reference \s{}-shifts. 
This also holds when neglecting differential polarization effects. 

The effect of differential polarization on the \f{}-shifts seems more pronounced than that of the \s{}-shifts
 in standard PE. 
This effect on the \f{}-shifts is, however, reduced by the incorporation of external effective field effects, so that for a 4~{\AA} environment of \pna{}, the average \f{}-shift is similarly well described with and without differential polarization. 
For individual snapshots, however, the effect of differential polarization can be sizeable, both with and without including external effective field effects. 
 In these cases, external field effects improve the agreement with the supersystem reference.
We could further show that the severe electron-spill-out issues preventing traditional PE calculations on the highly anionic pFTAA dye with sodium ions in close proximity could be largely reduced by atomic pseudopotentials on the sodium ions.  
  
All in all, we find a similar performance for FDE and PE on excitation energies as well as average oscillator strengths.
Accurate oscillator strengths with PE, however, required the incorporation of external effective field effects. 
For the anionic pFTAA example, further effective core potentials on nearby cations were essential to avoid electron-spill-out effects.  
  
\section*{Author contributions}
MJ: Methodology, Software, Validation, Formal Analysis, Investigation, Data Curation, Writing-Original Draft, Visualization;
PR: Molecular Dynamics for \pna{} in water;  
EDH: Conceptualization, Validation, Formal Analysis, Writing - Review \& Editing; 
CK: Conceptualization, Validation, Formal Analysis, Writing - Review \& Editing, Supervision, Project Administration.

\section*{Acknowledgments}
 This work has been supported by the Deutsche Forschungsgemeinschaft (DFG) through the Emmy Noether Young Group Leader Programme (project KO 5423/1-1). EDH thanks The Villum Foundation, Young Investigator Program (grant no.~29412), the
Swedish Research Council (grant no.~2019-04205), and Independent Research Fund Denmark (grant no.~0252-00002B and grant no.~2064-00002B) for support.

\printbibliography

\end{document}